\documentclass[12pt]{iopart}
\pdfoutput=1
\usepackage{iopams}
\usepackage{ulem}
\usepackage{graphicx}
\usepackage{color}
\usepackage{subfig}
\begin{document}

\title[Model for proportionate growth]{A sandpile model for proportionate growth}
\author{Deepak Dhar$^\S$ and Tridib Sadhu$^\dagger$ }
\address{$^\S$ Department of Theoretical Physics, Tata Institute of Fundamental
Research,\\ Homi
Bhabha Road, Mumbai 400 005, India.}
\address{$^\dagger$ Institut de Physique Théorique,
Orme des Merisiers,\\ batiment 774 CEA/DSM/IPhT, CEA/Saclay
F-91191, Gif-sur-Yvette Cedex, France.}
\eads{$^\S$ \mailto{ddhar@tifr.res.in},\quad$^\dagger$ \mailto{tridib.sadhu@cea.fr}}
\date{\today}
\begin{abstract}
An interesting feature of growth in animals is that different parts of the body
grow at approximately the same rate. This property is called proportionate growth.
In this paper, we review our recent work on patterns formed by adding $N$ grains
at a single site in the abelian sandpile model. These simple models show
very intricate patterns,  show proportionate growth, and sometimes having a striking resemblance to natural forms. We give several examples of such
patterns. We discuss the special cases where the asymptotic pattern can be determined
exactly. The effect of noise in the background or in the rules on the patterns is also
discussed.
\end{abstract}
\pacs{ }
\submitto{JSTAT}
\maketitle

\section{Introduction}
In this paper, we will review our recent work on  patterns formed by growing sandpiles.
The motivation for this study comes from different directions. Firstly, growing
sandpiles  provide a simple model of a well-known biological phenomena: as an
animal grows from birth to adulthood, different parts of the body grow at  roughly the
same rate.  Secondly, one can get a large variety of intricate beautiful
patterns, and these can be charaterized exactly, and thus provide a useful
theoretical model of pattern formation. And finally, the exact characterization
involves the application of some interesting mathematics, from the theory of
discrete analytic functions to tropical algebra.
This is a written version of the talk given at Statphys 25. A shorter, less
technical, review was prepared earlier \cite{tridib5}.

The plan of this paper is as follows. We discuss the basic phenomenology of
proportionate growth in \sref{sec:prop growth}.  In \sref{sec:soc}, we briefly discuss the
historical development of the ideas of self-organization and self-organized
criticality. \Sref{sec:model}
introduces the sandpile model, and gives some example of patterns that are
generated. \Sref{sec:scaling} develops the general scaling theory for describing the
patterns in terms of the scaled toppling function, and  we describe the basic
result about the piece-wise linear or quadratic nature of the scaled toppling
function in the periodic patches. This result forms the basis of exact
characterization of patterns that is sketched in \sref{sec:characterization}.
\Sref{sec:noise}
discusses the patterns formed on disordered initial substrates, or perturbed by
presence of boundaries etc. \Sref{sec:anisotropy} discusses the effect of introducing
dissipation or anisotropy in the toppling rules. \Sref{sec:conclusion} discusses the
striking similarity of some of the patterns generated with very simple rules
with natural once, and  possible directions for further work.

\section{Proportionate growth in animals \label{sec:prop growth}}
\begin{figure}
\begin{center}
\includegraphics[width=6.0cm]{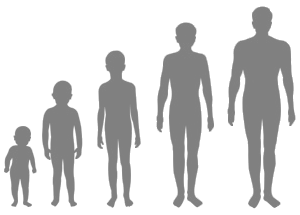}
\caption{Proportionate growth in humans.\label{fig:human growth}}
\end{center}
\end{figure}
Growth and development of different structures in animals and
plants has been a source
of fascination and bewilderment for scientists and laymen alike. As a baby
animal grows into an adult, it is often seen that different body parts in
animals grow roughly at the same rate, keeping the overall shape approximately unchanged
(see \fref{fig:human growth}). This property is called proportionate
growth. Of course, this is not exactly true, \textit{e.g.}, in humans, limbs grow
faster than head, some physical changes appear at puberty,
\textit{etc}. However, proportionate growth provides a good starting description. We would like to argue that in spite of a lot of work in
developmental biology for over a century, this basic phenomena in the problem
of development is not well-understood.

The usual biological explanation of this
would invoke different chemicals (hormones, growth factors...) that turn on and off
the production of other chemicals according to the genetic master plan given in
the animal's DNA. But this approach, focussed on precise identification of
chemical agents for various processes is not quite satisfactory. It is like
saying in a murder mystery, ``the knife did it''. We would like to look at this
problem from a physicist's perspective, and would
like to construct simple physics models that show proportionate
growth. Following the philosophy of d'Arcy Thompson \cite{d'arcy_thompson}, we would like to focus
on the interplay of growth and geometrical structures, without getting
bogged down in details of chemistry.  The models we study are qualitatively different from other models of growth that have been studied  by physicists in the past.  Typical examples of growing
structures that have been studied  in physics so far are growth of crystals from a super-saturated solution, or
diffusion limited aggregation, or viscous fingering, surface growth in
molecular beam epitaxy. In all these cases, the growth occurs on the outer
surface, and the inner parts once formed remain frozen. In fact, finding
systems that show proportionate growth outside biology is rather difficult.  
This is because proportionate growth implies regulation, which requires
some communication and long-range interaction between different parts of the
body: this  is not easily captured in the simpler processes mentioned above.

\section{Self-organization and sandpiles \label{sec:soc}}
In the 1970's, Haken, Nicolis, Prigogine and coworkers emphasized that an important characteristic of  living systems is that they are   `self-organized' \cite{nicolis,haken}. Here  `self-organized' is
not just an autonomous system, but a non-equilibrium steady state that has some
internal self-regulation, and typically shows, amongst other things, 
self-generated complex spatial structures.

In 1987, Bak \etal extended this idea of self-organization to other classes of
natural systems out of equilibrium, like earthquakes and solar flares. They
noted that these systems by their own natural dynamics tend to, and stay at, a critical state at the edge of stability, and called these {\it Self-Organized Critical} \cite{btw87}. They
proposed a simple model of sandpiles to illustrate this idea. A pile formed by
dropping sand on a flat table is critical in the sense that the relaxation
event of the
system on the addition of a single grain (called an avalanche), has a wide
distribution of sizes. This model has an interesting mathematical structure, and
has inspired a large number of studies, of this, and other models of
self-organized criticality \cite{bak_book,dd2006}.
However, our interest in this paper is not in the criticality shown by the
sandpile model at its  steady state, but \textit{the self-organization} shown in the
interesting pattern that are generated in \textit{growing sandpiles}.
We hope that study of growing patterns generated in the sandpile model will also
help in a better understanding of the original questions that led to
its study.

We will show in this paper that the abelian sandpile model provides a very
interesting model of pattern formation and proportionate growth. It is
analytically tractable, and at least in some simpler cases, the asymptotic
patterns can be fully characterized. It thus adds to the small number of known
analytically tractable complex systems with simple rules.

\begin{figure}
	\begin{center}
\includegraphics[width=5.0cm]{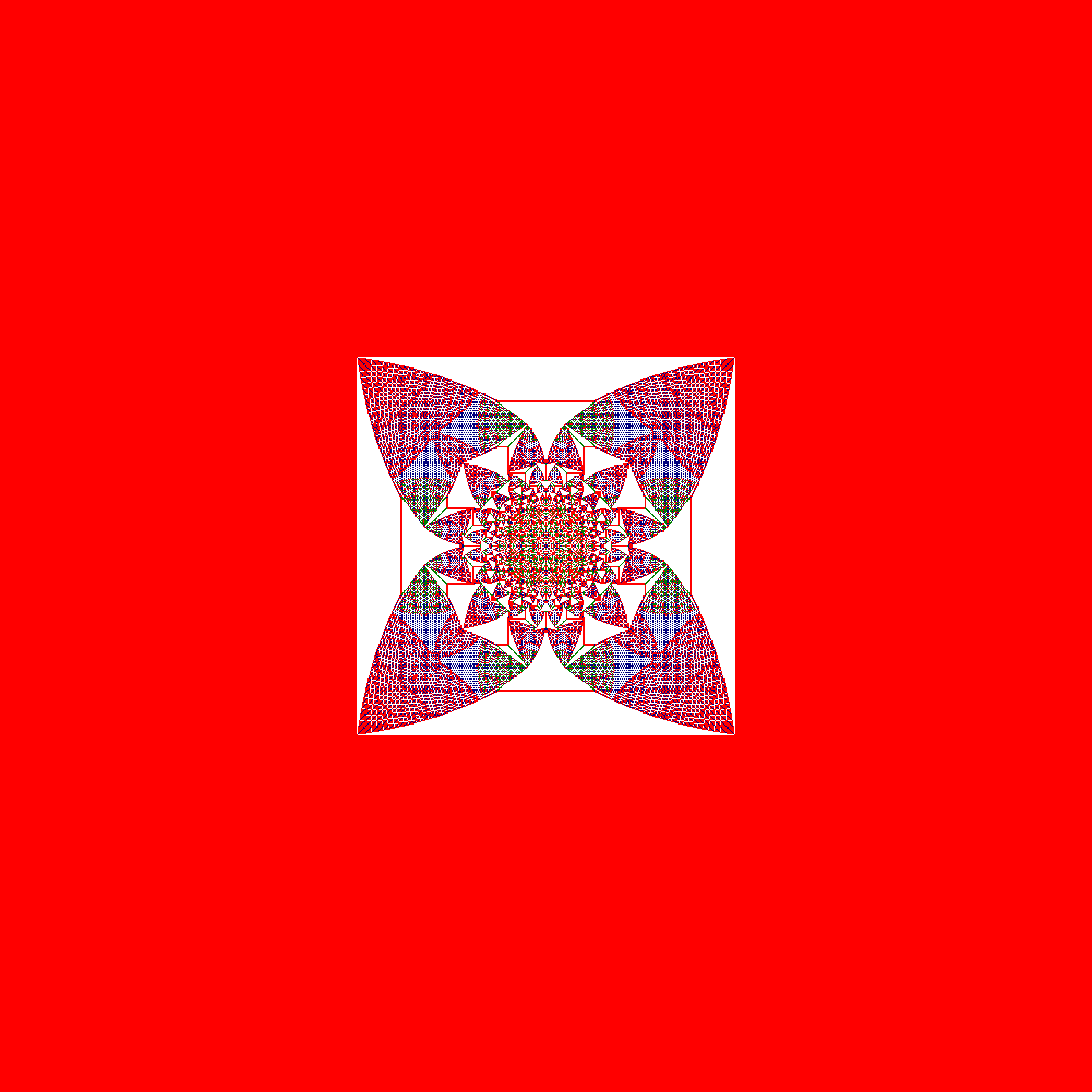}
\includegraphics[width=5.0cm]{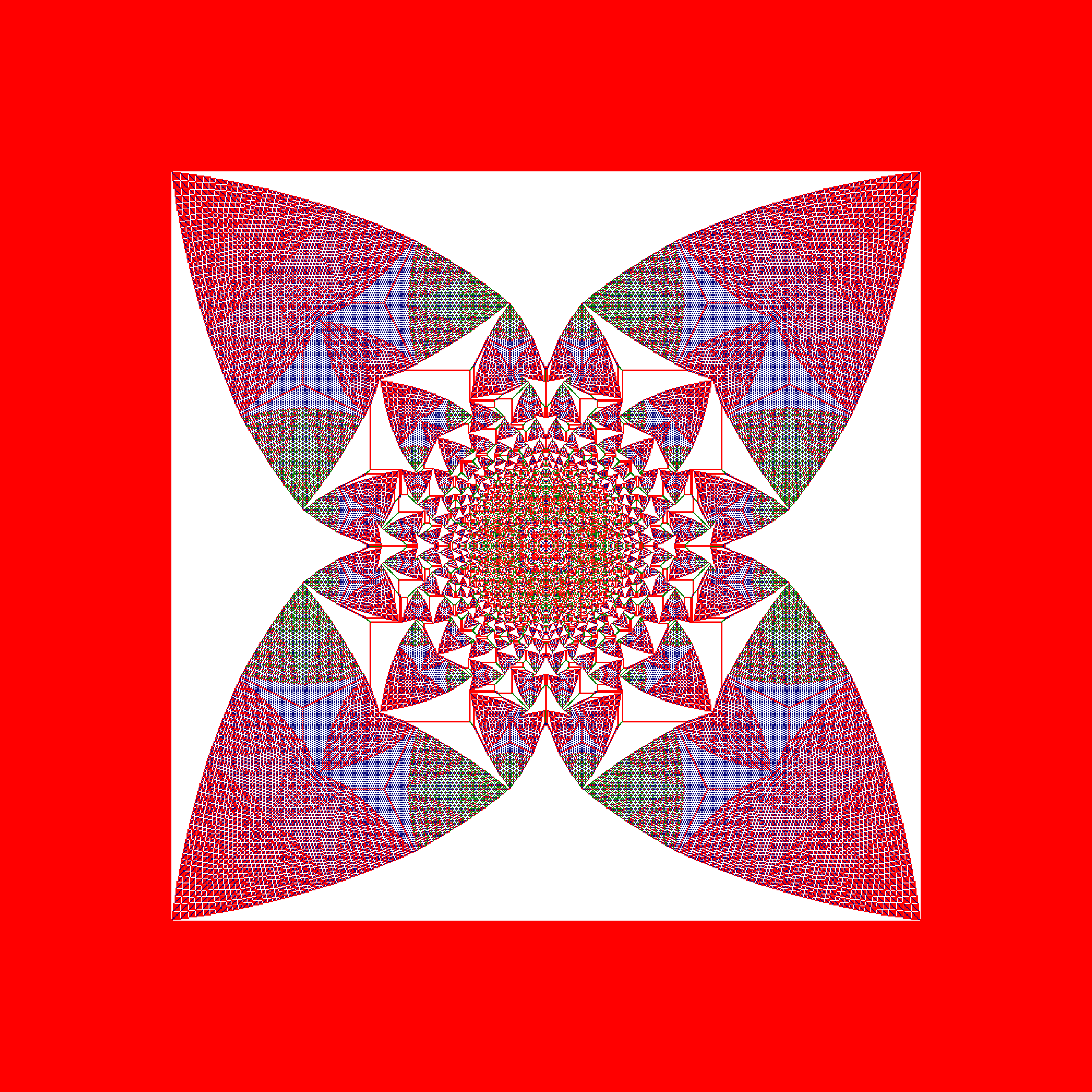}
\includegraphics[width=5.0cm]{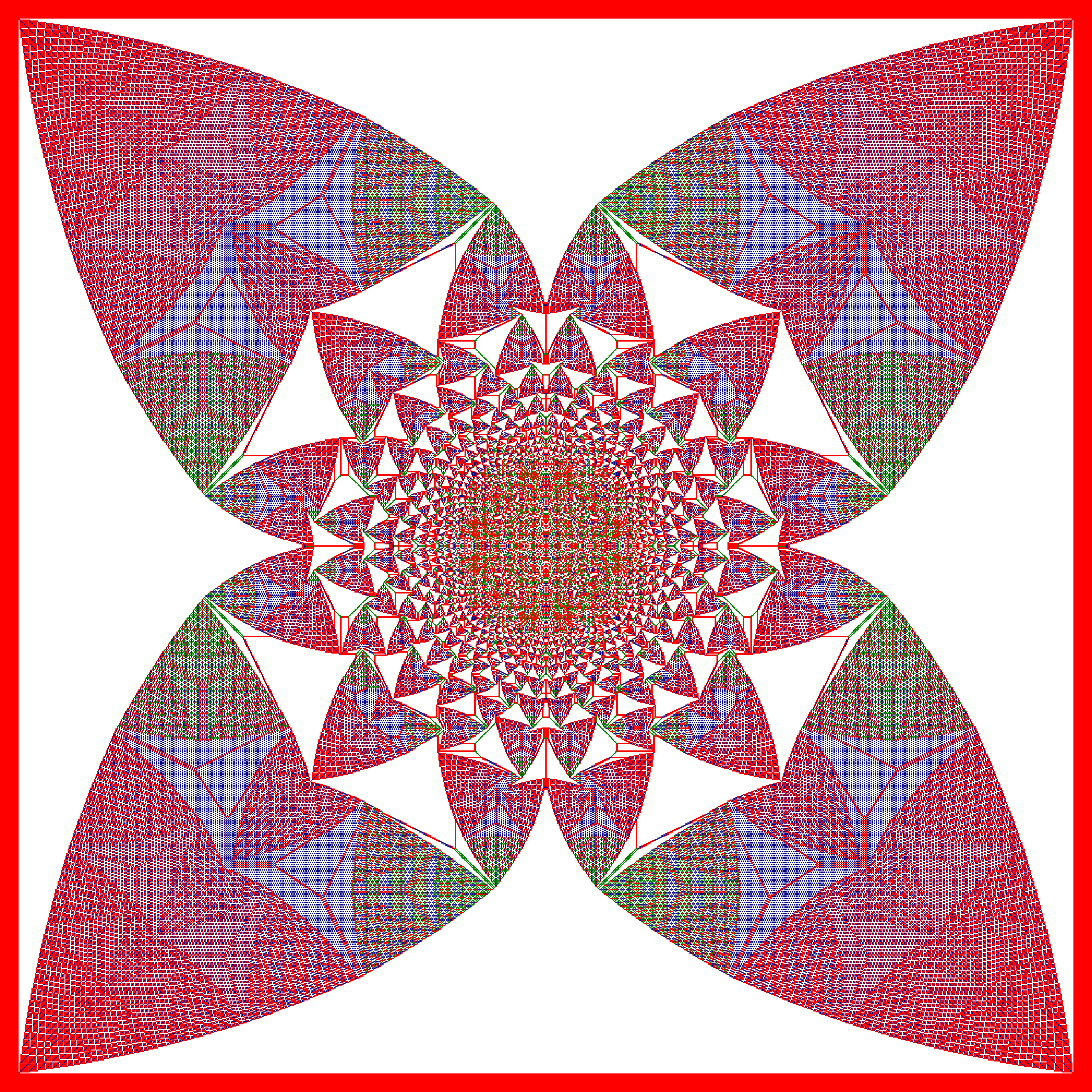}
\caption{Patterns formed on a square lattice with initial height 2 at all sites for $N=  4\times10^4$, $2\times10^5$, and $4\times10^5$.
Color code: $0, 1, 2, 3=$ B, G, R, W. The patterns are on
the same scale. Zoom in for details in the electronic version. Figures reproduced from
\cite{tridib5}. \label{fig:proportionate}}
	\end{center}
\end{figure}

\begin{figure}
\begin{center}
	\fbox{\includegraphics[width=3.5cm]{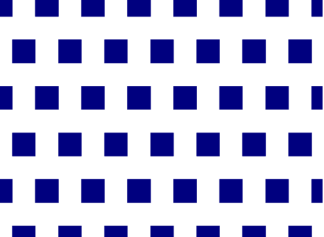}}
\qquad
\fbox{\includegraphics[width=3.6cm]{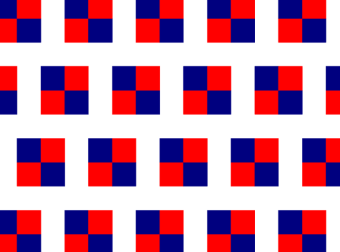}}
\qquad
\fbox{\includegraphics[width=3.6cm]{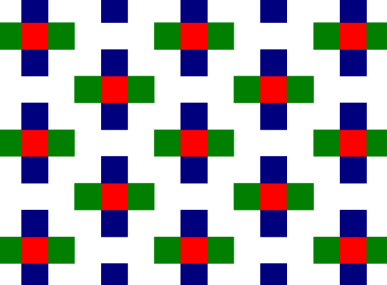}}
\caption{Some examples of periodic patterns in heights in
\fref{fig:proportionate}, obtained by zooming into different
patches. Color code same as in the original figure. Figures reproduced from
\cite{tridib5}. \label{fig:patches}}
\end{center}
\end{figure}

\section{Definition of the model \label{sec:model}}
To make a primitive  and simple model of biological growth, we note only some
basic facts from biology. The first is that food is required for
growth. Intake of
food is typically from a localized organ (the `mouth'), and
the nutrients are
then transported to all parts of the body. The second is that the basic
process in biological growth is cell-division. This is a threshold process in
the sense that a cell will not divide if it does not have enough nutrients. And the third is that   same food
becomes different tissues in different parts of the body. 

A well-studied  model of threshold dynamics is the Abelian Sandpile Model
\cite{btw87,dd90}. For simplicity, we define it on a square lattice. At each
site $i$ there is a non-negative integer height variable $z_i$, which is called the
number of sand grains at $i$. We say that a site is unstable, if the height at
the site exceeds $3$. An unstable site relaxes by toppling: the height at the
site is decreased by $4$, and the height at each neighboring site is increased
by $1$. If this results in any of the neighbors becoming unstable, they are
relaxed in the same way, until all sites are stable.
\begin{figure}
\begin{center}
\includegraphics[width=7.0cm]{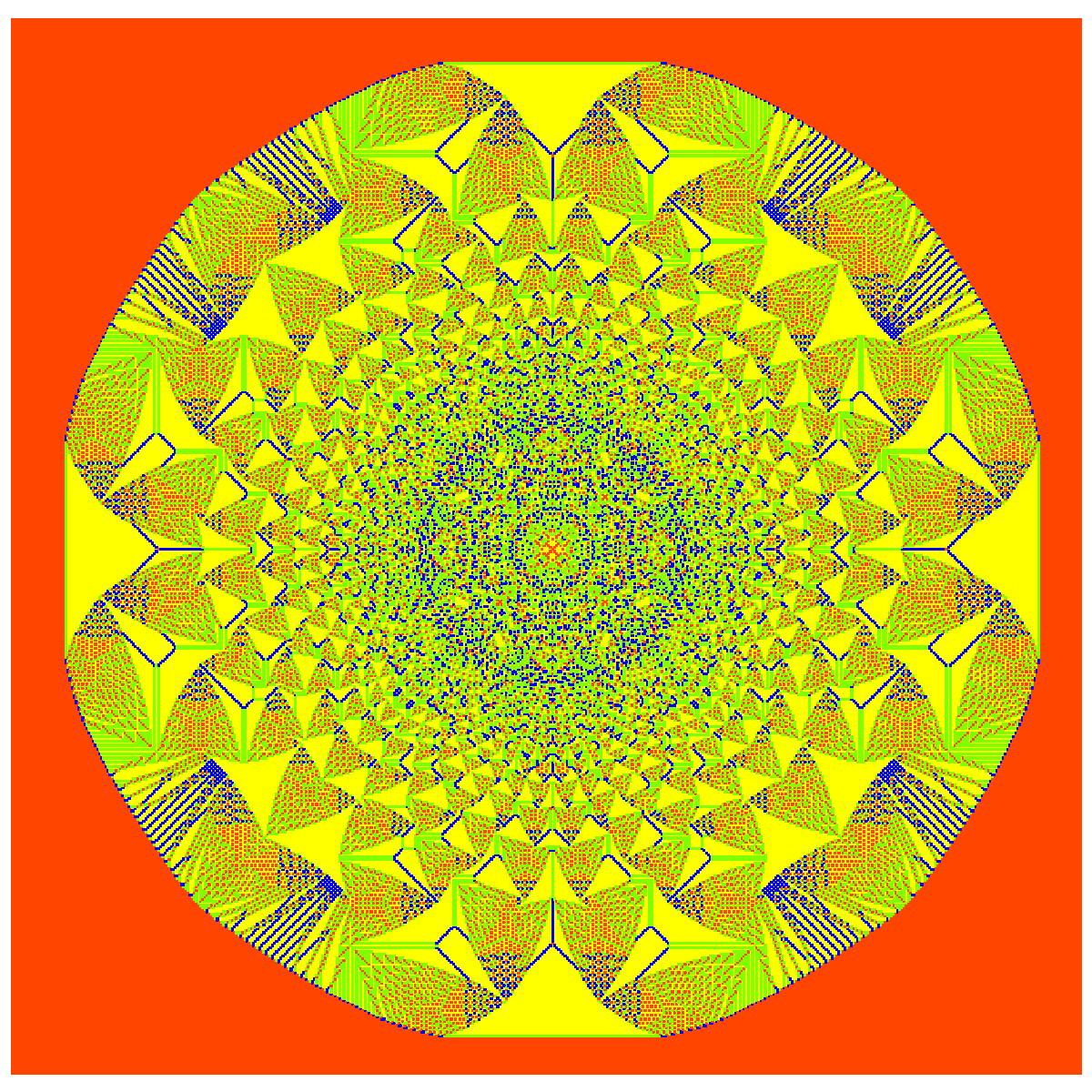}
\caption{Pattern produced by adding $N=4\times 10^5$ particles at the origin, on a square
lattice ASM, with initial state of all height $0$. Color code: $0$, $1$, $2$, $3 =$
R, B, G, Y. Zoom in for details in the electronic version. \label{fig:allzero}}
\end{center}
\end{figure}

We start with a periodic stable configuration of heights on an infinite lattice.
Then, we add one grain at the origin, and relax the configuration, if unstable.
Then add another grain, and relax. And so on.

This model has the property that if a configuration has several unstable sites,
the order in which they are relaxed does not matter. The operations of
adding particles at different sites and relaxing commute with each other. The
operators form an abelian group, and hence this model is called the abelian
sandpile model. The pattern obtained after $N$ grains have been added is a
deterministic pattern. For different starting backgrounds, one gets different
patterns.

In \fref{fig:proportionate}, we have shown the patterns corresponding to three different values of $N$, starting with a
background with all sites with height $2$. We see that the pattern for larger
$N$ is bigger, but has a similar structure: it shows proportionate growth.
It is important to note that as pattern grows, structures at finer length scale
are formed near the origin. Once formed, they grow proportionately and move away
from origin.

The pattern in \fref{fig:proportionate} consists of distinct structures, called patches here, with sharp
boundaries. The very interesting and unexpected observation is that within a
patch, the arrangement of heights forms a perfectly periodic structure, with
only a few `defect lines'. Some examples of the periodic structures are shown in
\fref{fig:patches}. When $N$ is increased, the sizes of the patches increase,
and their location on the lattice will also change. The whole pattern consists
of these periodic patches sewn together into a quilt-like whole.

One can similarly define the sandpile model on other lattices, and study
patterns in other backgrounds. In higher dimensions also, we see the same
phenomena, but we will confine our discussion
here mostly to two dimensions, for ease of display. In \fref{fig:allzero}, we show the pattern on the
square lattice, when the background is all sites having height zero. In this case
also, we see proportionate growth.

\begin{figure}
\begin{center}
	\centering
\includegraphics[width=5.0cm]{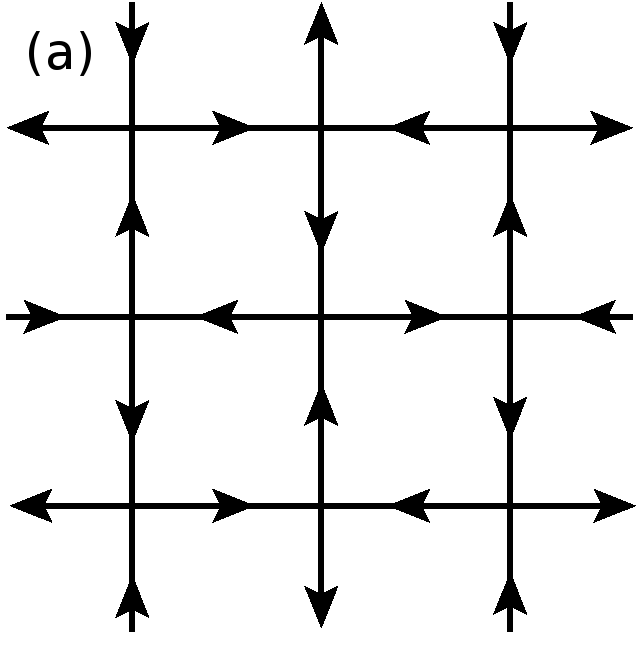}
\qquad
\includegraphics[width=6.0cm]{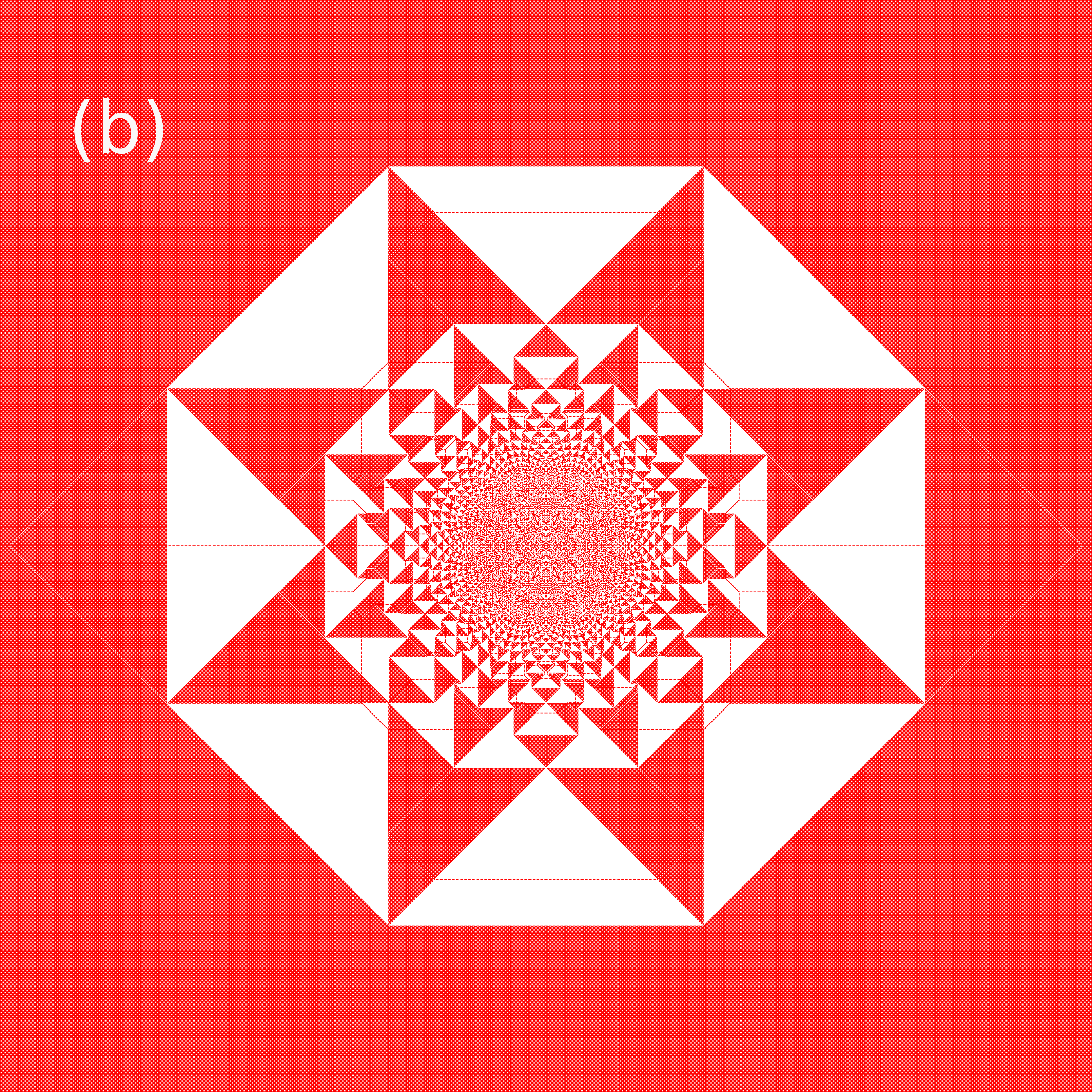}
\caption{(a) The F-lattice, with two in-arrows and two out-arrows at each
vertex of a square lattice. (b) The pattern produced by adding $2\times10^5$ particles
at the origin, on the F-lattice with initial background being checkerboard.
Color code: $0=$ red, $1=$ white. The apparent red coloured patch is actually a
checkerboard arrangement of red and white. Zoom in for details in the electronic version. \label{fig:octagon}}
\end{center}
\end{figure}

In \fref{fig:octagon}(a), we show the graph defining the F-lattice. This is
a directed square lattice with two arrows in and two arrows out at each vertex.
The sandpile model on this graph is defined by the rule that any site with
number of grains $\geq 2$ is unstable, and topples by sending two grains along
the  outward arrows. If we start with a checkerboard background with alternate
sites having $1$ and $0$, and then add $N$ grains at the origin, the resultant
pattern is shown in \fref{fig:octagon}(b). In particular, we note here that the asymptotic pattern shows an
eight-fold rotational symmetry, higher than the symmetry of the underlying
lattice.

Patterns produced in sandpile models have been studied almost as long as
sandpiles themselves. Liu \etal \cite{liu} noted that
patterns produced by relaxing special unstable states in the abelian sandpile
model have complex fractal-like internal structures.
First studies dealt with the shape of the boundary of the toppled cluster in a centrally seeded sandpile \cite{dhar99}. Bounds on the rate of growth of these boundaries were
obtained in \cite{borgne} and subsequently improved by \cite{redig,lionel2}. First detailed study of the periodic structures
in the pattern was by Ostojic \cite{srdjan}. He also noted the
piece-wise quadratic nature of the toppling functions within a patch. The first pattern to be 
characterized fully was  a
sandpile pattern on the F lattice \cite{tridib1}.  In this case, the asyptotic pattern was determined in terms of the solution  of the Laplace equation on the adjacency graph of the pattern.  This technique was subsequently  used in \cite{tridib2} to characterize
patterns with more than one site of addition and also those produced in presence of
absorbing sites. The effect of external noise on the pattern were studied in \cite{tridib4}.
We have not been able to characterize fully the pattern on the square lattice in
a similar way. In this case when the background is with all heights zero, the
existence of a limit pattern has been proved rigorously \cite{pegden}.

There are other spatial patterns formed in the sandpile model, like the identity
\cite{borgne,creutz} or the
configurations produced by relaxing from a uniform unstable state
\cite{liu}. These also show complex self-similar structures which are similar to
those studied in this paper \cite{sportiello}.

Growing complex patterns has also been studied in other models, similar to the sandpile model.
For example, in the Internal Diffusion Limited Aggregation model \cite{lawler} it has been shown
that the boundary of the asymptotic pattern is related to the classical Stefan
problem in hydrodynamics \cite{gravner}. Levine and Peres have proved the
existence of a limit shape for the pattern with multiple sources \cite{lionel3}. There are
several other studies on patterns in Eulerian walkers
\cite{abhishek,lionel1,propprotor,rahul}, infinitely divisible
sandpiles and non-Abelian sandpiles \cite{nonab}.  

\section{General scaling theory for the asymptotic patterns \label{sec:scaling}} 
Let us denote the diameter of the pattern for a given value of $N$ by $\Lambda$.
We define a reduced coordinate $\vec{r} \equiv(x,y)= \vec{R}/
\Lambda\equiv(X/\Lambda,Y/\Lambda)$. Let  $ T_N
(\vec{R})$ be the number of topplings at point $\vec{R}$. Then, the property of
proportionate growth may be stated in terms of the scaling of $T_N (\vec{R}) $
with $N$, or equivalently, with  $\Lambda$. If the pattern shows proportionate
growth, then, up to an overall multiplicative constant, the leading behavior of
$T_N$ depends only on the reduced coordinates $\vec{r}$:
\begin{equation}
	T_N (\vec{R}) \sim \Lambda^a \phi(\vec{R}/\Lambda).
\end{equation}

In addition, we expect that $\Lambda \sim N^{1/b}$, where $a$ and $b$ are some exponents.
More formally, we define the function $\phi(\vec{r})$ by the equation

\begin{equation}
\phi(x,y) = \lim_{\Lambda \rightarrow \infty} \Lambda^{-a} T_N( \lfloor \Lambda
x\rfloor,\lfloor \Lambda y\rfloor),
\end{equation}
where $\lfloor x \rfloor$ is the largest integer less than or equal to
$x$. If this limit exists, with a non-trivial function $\phi(\vec{r})$, then the
pattern shows proportionate growth.
\begin{figure}
\begin{center}
\includegraphics[width=7.0cm]{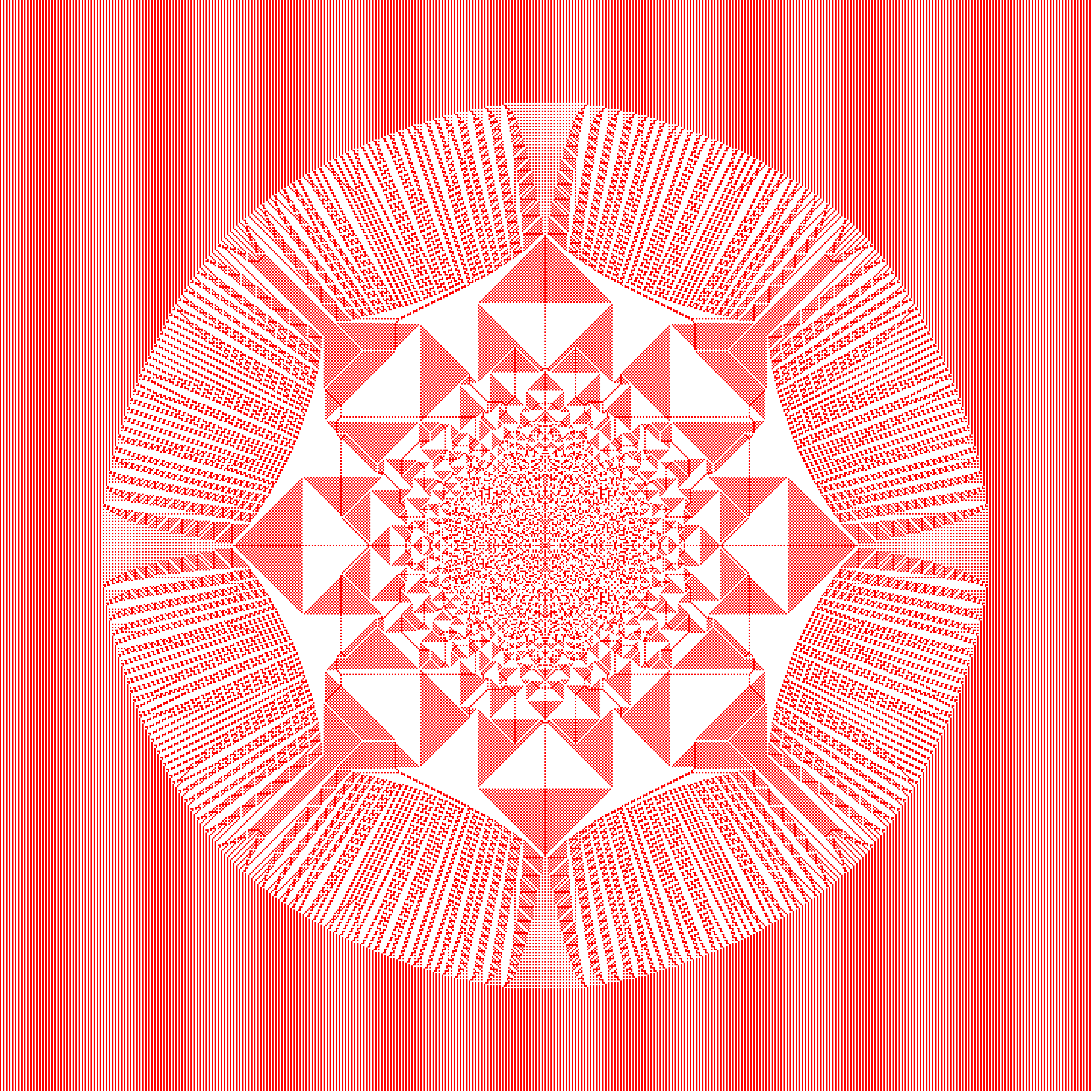}
\caption{An example of a pattern with patches with aperiodic structures. This
is produced by adding $N=10^4$ particles at the origin, on the F-lattice ASM,
with initial state of alternating columns of $1$'s and $0$'s. Color code: red
$=0$, white $=1$. Zoom in for details in the electronic version.
\label{fig:non-periodic}}
\end{center}
\end{figure}

We note that lattice laplacian of $T_N(
X,Y)$ gives the local change in height, and hence specifies the asymptotic
height pattern. The function $\nabla^2 \phi (x,y)$ gives the local density of
excess particles in the neighborhood of a point corresponding to the reduced
coordinate $(x,y)$.
The excess number of grains per site  is bounded everywhere. This implies that
\begin{equation}
a \leq 2.
\end{equation}

Our analysis of the sandpile patterns depends on the following remarkable
theorem: in each patch with a periodic height pattern, we can only have
$a =2$ or $a=1$. In addition,  within a periodic patch, $\phi(x,y)$ is a polynomial function of $x$ and $y$ of degree $a$. 

The proof depends on the fact that $T_N(X,Y)$ is an integer function of  integer
arguments $X$ and $Y$, and any higher order terms in the Taylor expansion of
$\phi(\vec{r})$ are not consistent with this condition.  For example, a cubic
term of the form $ K (\Delta x)^3$ in the Taylor expansion of $\phi(x_0 + \Delta
x, y_0 + \Delta y)$  can only come from a term of the form $ K (\Delta X)^3/
\Lambda^{3-a}$ in the expansion of $T_N$.  But given that $\Delta X$ and $T_N$
are both integers, this would require defect lines with a spacing of order
$\Lambda^{1-a/3}$. Since a periodic patch which itself has diameter of order
$\Lambda$, by definition, does not have any defect
structures having this intermediate scale of length, we conclude that $K=0$. Similar
argument holds for other cubic, or higher order terms in the Taylor expansion.
For details, see \cite{tridib6}.

Sometimes, the generated patterns does show a large number of defect lines. Examples
are the patterns in \fref{fig:allzero} and
\fref{fig:non-periodic}. The  latter pattern is  generated on the F-lattice,
when the initial background consisted of alternate columns of 1's and zeroes.
The inner part of the pattern consists of periodic patches, where each patch
occupies a non-zero fraction of the area of the pattern. However, in the outer
rim, we see a large number of radial defect lines.  Thus, for this pattern, it
would appear that the function $\phi(\vec{r})$ is  a piece-wise quadratic
function of $\vec{r}$ in the periodic patches, {\it except in the outer rim
region}.  Of course, it may be that the outer rim region, is not a single large
patch, but a union of many periodic patches.

Interestingly, the value of the exponent $b$ is much less constrained, and we can
get different values of $b$, with $ b \le d$
($d$ is the dimension of the lattice), depending on the initial
periodic background chosen.

If the density of particles is low enough, then one gets a pattern where
$\Lambda \sim N^{1/d}$, where $d$ is the dimension of the lattice. For example,
for the $d$-dimensional hypercubic lattice, if all sites in the initial pattern
have heights $\leq (2d-2)$, where the threshold height is $2d$, then the pattern has $\Lambda \sim N^{1/d}$. If on
the other hand, sufficiently many sites have height $(2d-1)$, \textit{e.g.}, if they form a
spanning cluster, then clearly, we get infinite avalanches, and $\Lambda $ is
infinite for finite $N$.

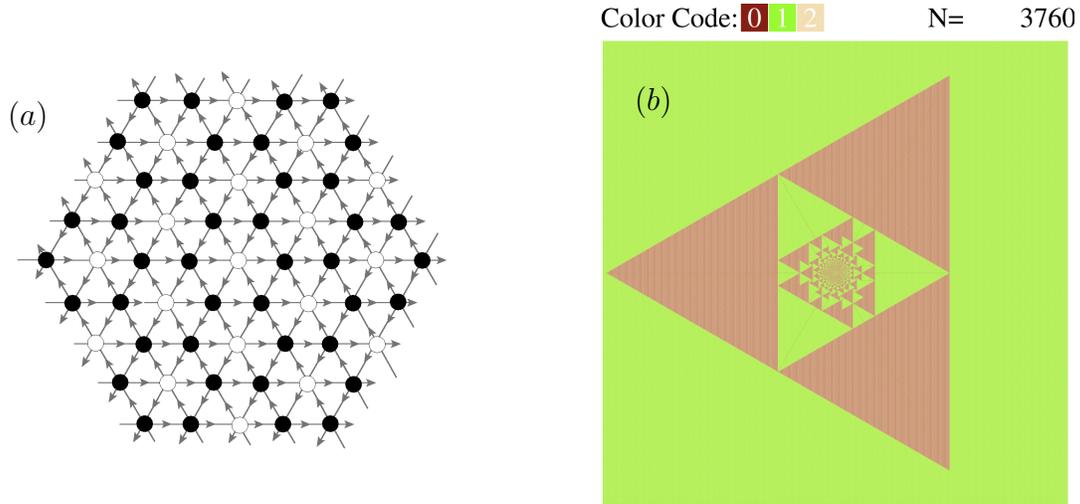
\begin{figure}
	\vskip-5ex
	\subfloat{
	\begin{minipage}[c][1\width]{0.5\textwidth}
		\centering
		\includegraphics[width=0.8\textwidth]{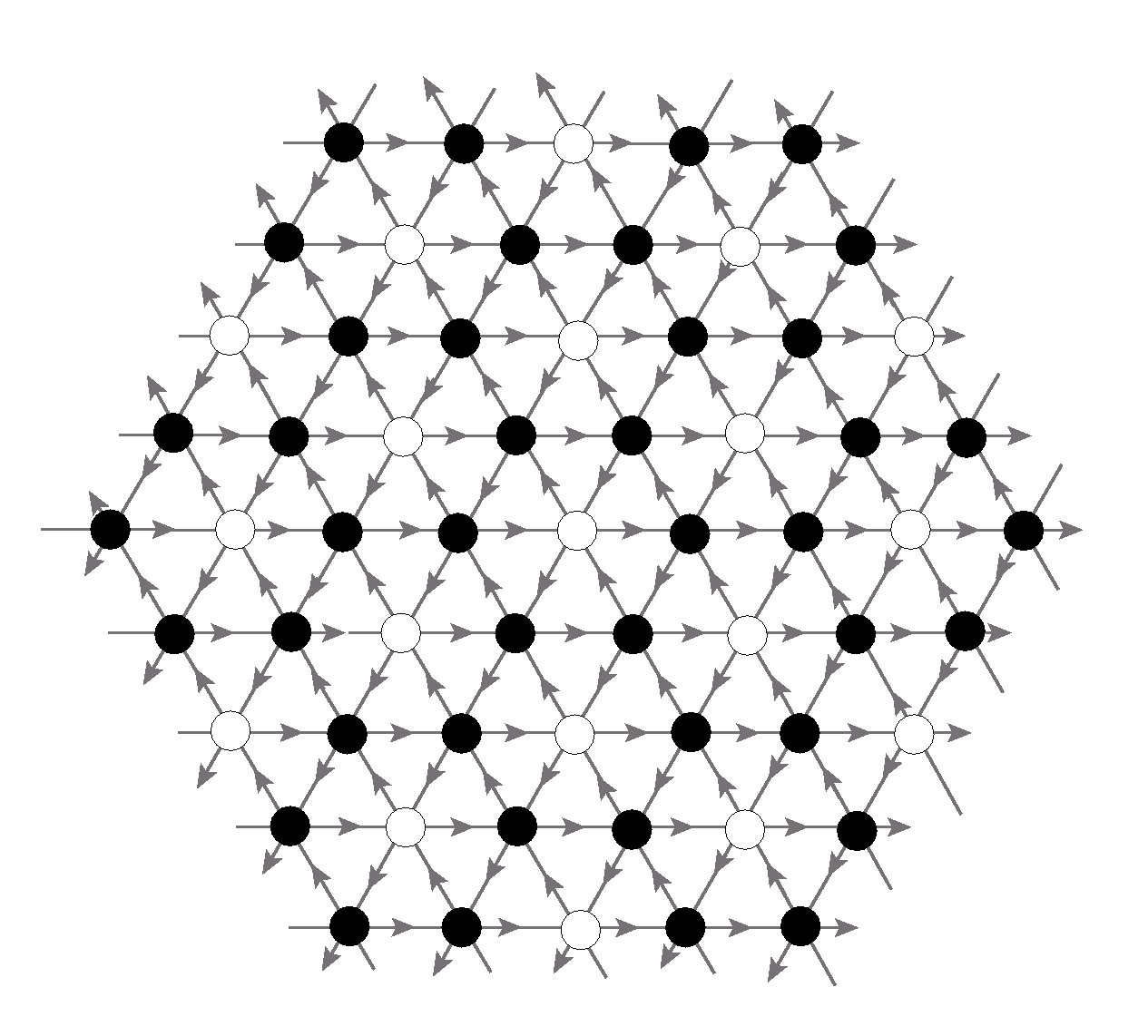}
\begin{picture}(0,0)
	\put(-180,130){$(a)$}
\end{picture}
	\end{minipage}}
	\subfloat{
	\begin{minipage}[c][1\width]{0.5\textwidth}
		\centering
	\includegraphics[width=0.8\textwidth]{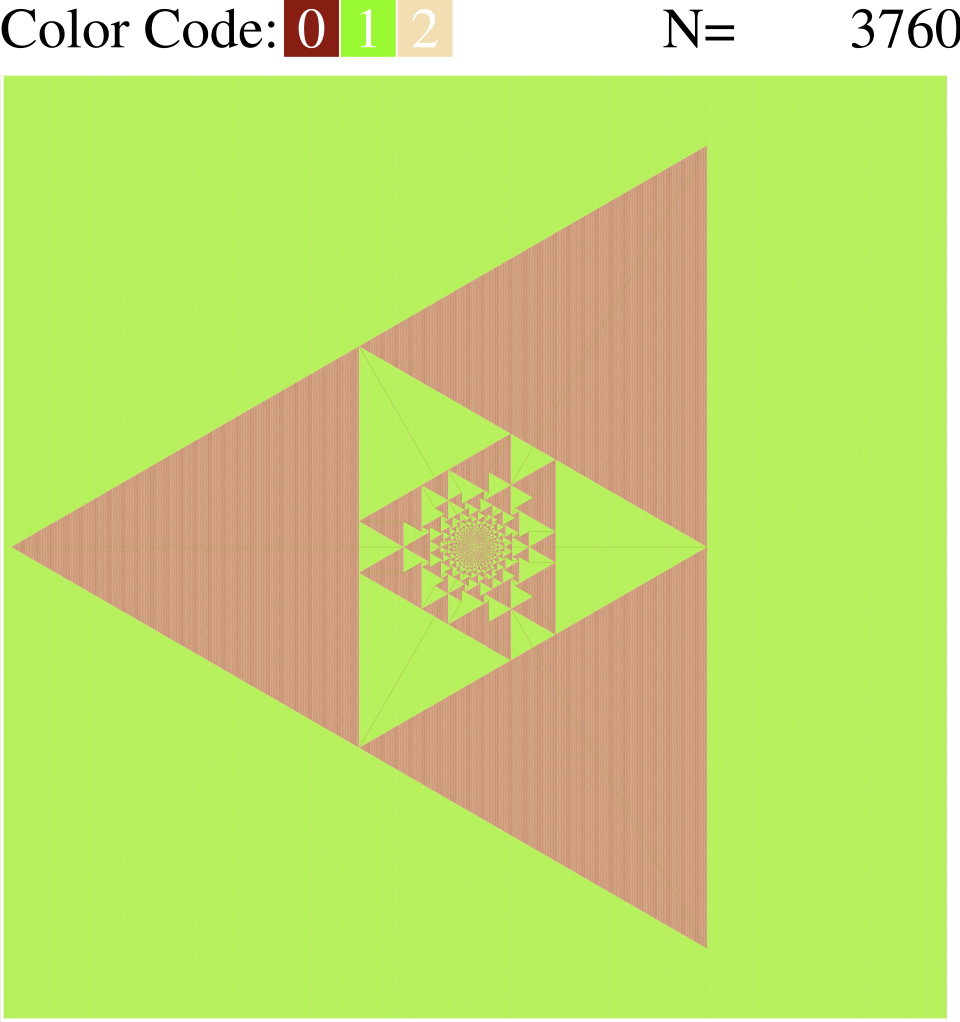}
\begin{picture}(0,0)
	\put(-170,150){$(b)$}
\end{picture}
	\end{minipage}}
\caption{(a) Directed triangular lattice. In the background configuration, filled
and unfilled circles denote $z=1$ and $2$. (b) The pattern produced on the
background by adding $N=3760$ particles. The apparent green patch denotes
background height configuration. Zoom in for details in the electronic version.
Reproduced from \cite{tridib6}.
\label{fig:triangle}}
\end{figure}

In \fref{fig:triangle}, we have shown a directed triangular lattice, with
three arrows in, and three out of each vertex. The toppling rule for the sandpile is
that any site with more than $2$ grains is unstable, and transfers one grain
each in the direction of outgoing arrows. Starting with the background shown in
\fref{fig:triangle}(a), the pattern generated by adding grains and relaxing is
shown in \fref{fig:triangle}(b). In this case, $\Lambda \sim N$, and $b
=1$.  In \cite{tridib6}, we have discussed an infinite family  of initial
backgrounds on this lattice  that all give $b=1$.

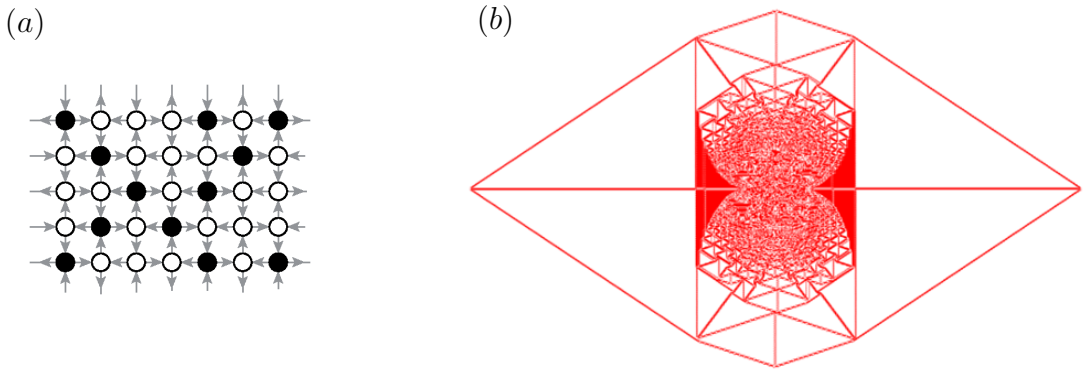
\begin{figure}
	\vskip-18ex
	\subfloat{
	\begin{minipage}[c][1\width]{0.3\textwidth}
		\centering
		\includegraphics[width=0.8\textwidth]{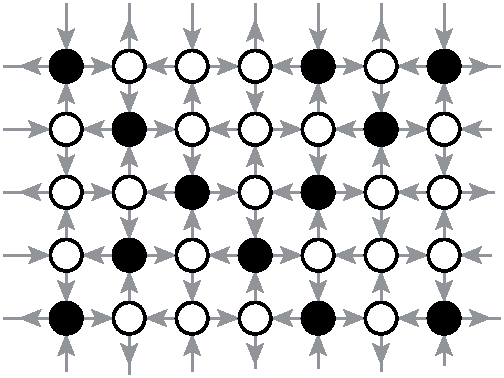}
\begin{picture}(0,0)
	\put(-120,100){$(a)$}
\end{picture}
	\end{minipage}}
	\subfloat{
	\begin{minipage}[c][1\width]{0.7\textwidth}
		\centering
	\includegraphics[width=0.8\textwidth]{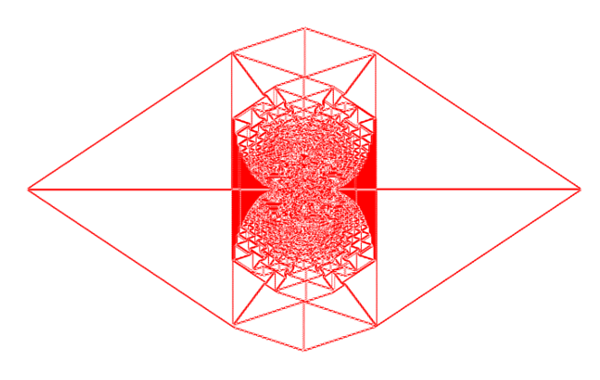}
\begin{picture}(0,0)
	\put(-240,140){$(b)$}
\end{picture}
	\end{minipage}}
	\vskip-15ex
\caption{The `bat-pattern':  (a) the unit cell of the periodic initial pattern,
and (b) the pattern generated. Only the boundaries of patches are shown.
\label{fig:bat}}
\end{figure}

More interestingly, on the F-lattice, we found a family of initial backgrounds
that gives $b$ which lies between $2$ and $1$. In \fref{fig:bat} we show
the periodic background on the F-lattice, and the generated pattern.  When $b <
d$, the mean excess density in the patch in the asymptotic pattern is zero.  In
this case, the density inside the patches is same as in the background, and
added particles sit on the boundaries of patches. Some boundaries can also have a deficit of particles.  We have shown only the
boundaries of the patches. We will refer to this pattern as the `bat-pattern'. Our
numerical studies \cite{tridib6} have found that the wing-span of this bat
varies as $N^{b}$, with $b \approx 0.55$.  However, one also sees some region
with non-zero excess areal charge density in the pattern (solid colour in the figure).
The boundaries of these solid coloured regions seem to be exact parabolas, and so the width
of these regions would have to scale as $\Lambda^{1/2}$. This implies that
these solid coloured regions will shrink to two vertical lines in the asymptotic
pattern. The potential function would not be continuous along these lines.

There are also other periodic backgrounds on the F-lattice, for which we found other
values of $b$. The general shape is similar to that of the bat-pattern. For details, see
\cite{tridib6}.

\section{Exact characterization of the  patterns \label{sec:characterization}}
It is not immediately obvious how one can characterize these complex patterns in
detail. The first level of description  is clearly  structural. One describes
the patterns by describing which periodic patterns are found in which patch, and which patch is
adjacent to which. Formally, we define an adjacency graph for the pattern, where
the vertices are patches, and then we draw a link between two vertices if they
are adjacent. The pattern is described by giving its adjacency graph, and the
periodic pattern associated to each vertex. The next level of description gives
the metric properties of the patches: their position and exact equations of the
boundaries, \textit{etc}.

For the patterns that we have been able to characterize, we construct  the
adjacency graph by looking at the pattern.  It is easy to see the adjacency
relation between the larger patches. But the patches become smaller and more
numerous as we go closer to the origin, and they are not clearly resolved. In this
case, we extrapolate that the observed regularity of  structure will continue to hold
at higher resolutions.

The procedure is best illustrated by an example. To characterize the pattern in
\fref{fig:triangle}, we note that in this case $a=1$, and hence the potential
function is piece-wise linear.

\begin{enumerate}
	\item
		In a particular patch,  say denoted by $P$,  we will express the
		piece-wise linear function  $\phi(x,y)$ as
		\begin{equation}
			\phi_{P}(x,y)= a_P x +  b_P y + c_P.
		\end{equation}
		Here $a_P, b_P$ and $c_P$ are parameters that may be used to
		specify the patch. Since $T_N$ has to be an integer function, this
		implies that $a_p$ and $b_p$ are rational numbers. Different
		patches may then be represented as points in a two-dimensional
                plane with Euclidean coordinates $(a,b)$.               
		By actually examining the values of these parameters for several
		patches, we noticed that the allowed values of $(a_P,b_P)$ form a
		hexagonal lattice in this space, and patches corresponding
		to nearest neighbour vertices on this lattice are adjacent
		patches in the original pattern. In addition there are some
		additional adjacencies for patches lying along the six symmetry
		directions in the original pattern. Thus, each patch may be labelled by two
		integers $(\ell,m)$, which give the coordinates on the hexagonal
		lattice.
	\item
		The potential $\phi(x,y)$ is a continuous function of its
		arguments. Then, the equation of the boundary between two
		adjacent patches $P$ and $P'$ is obtained by the condition that
		$\phi_P = \phi_{P'}$ along the boundary.  From the linearity of
		$\phi_P$ and $\phi_{P'}$, it follows that all boundaries are
		straight lines.
	\item
		The condition that three patches $P, P'$ and $P''$ meet at a
		point implies a condition on the coefficients $c_P, c_{P'}$ and
		$c_{P''}$. It is easy to check that this condition implies that
		$c_{P}$ satisfies a Laplace equation on the hexagonal lattice in
		the adjacency graph (leaving out the extra edges in the graph that are not
		nearest neighbor edges on the hexagonal lattice).
	\item
		By solving the Laplace's equation on the infinite hexagonal
		lattice, we determine all the $c_P$'s.
\end{enumerate}

Once the potential in all the  patches is determined, we can find the position
of all patch boundaries, and the reduced coordinates  of all corners of patches in
the asymptotic pattern are determined exactly. For example, in this case, we find
that the equation  of the right boundary of the  triangle formed is $x = 1/3$.
Other backgrounds with similar adjacency structure, but slightly different
relative  sizes of patches are discussed in \cite{tridib6}.

For the patterns with $a=2$, the potential is a quadratic function in each
patch. The pattern in \fref{fig:octagon} is one such example. The adjacency graph has the
structure of a square grid on a two-sheeted Riemann surface. One needs six
coefficients to specify the quadratic function in each patch. Out of these, the three
coefficients of the quadratic terms are specified in terms of the integer coordinates $(l,m)$
of the patch on the adjacency graph. The coefficients of the linear term in $x$ and $y$ again are found
to satisfy the Laplace equation, and hence the solutions are determined in terms
of lattice Green function for the square grid on a two-sheeted Riemann surface. For details, see
\cite{tridib1}.

\section{Effect of noise \label{sec:noise}}
Clearly, deterministic cellular automaton models are not very realistic model of
biological growth, as real growth involves a fair amount of noise.  We  have
studied the effect of noise in several ways \cite{tridib4}. 

If the point where particles are added is random, but lies in a box of size $n$,
then when the diameter of the pattern is much bigger than $n$, the input acts
like a point source, and the asymptotic pattern is unchanged. If the region
where particles are added is a small region near the origin, but this region
also grows proportionately with the pattern, the intricate substructures of
patches in this small region is washed out, but the outer larger patches  have
the same appearance.

We also studied the effect of boundaries on the growing pattern. We considered
growth in a half-space, where only the points $(x,y)$ with $y \geq 0$ are available
for growth, and the  point of addition is the origin.  Any particle that is
transferred to a site outside the lattice is lost. In this case, a straight
forward analysis using the lattice propagator in the presence of a boundary
shows that the  diameter of the pattern grows as $N^{1/3}$. More generally, if the
growth occurs in a wedge of angle $\theta$, the diameter of the pattern grows as
$N^{\frac{\theta}{2 \pi + \theta}}$.
\begin{figure}
	\begin{center}
\includegraphics[width=6cm]{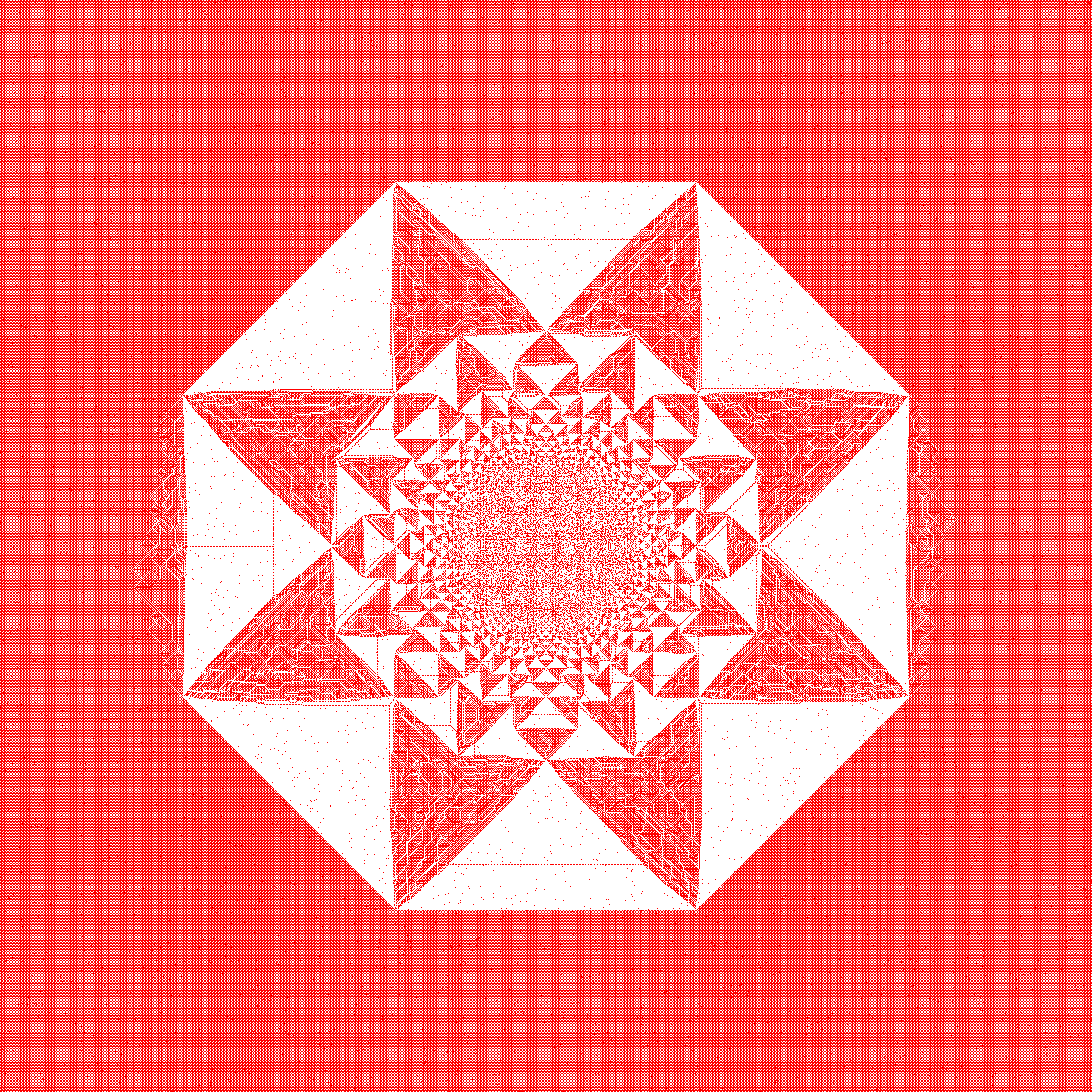}
\qquad
\includegraphics[width=6cm]{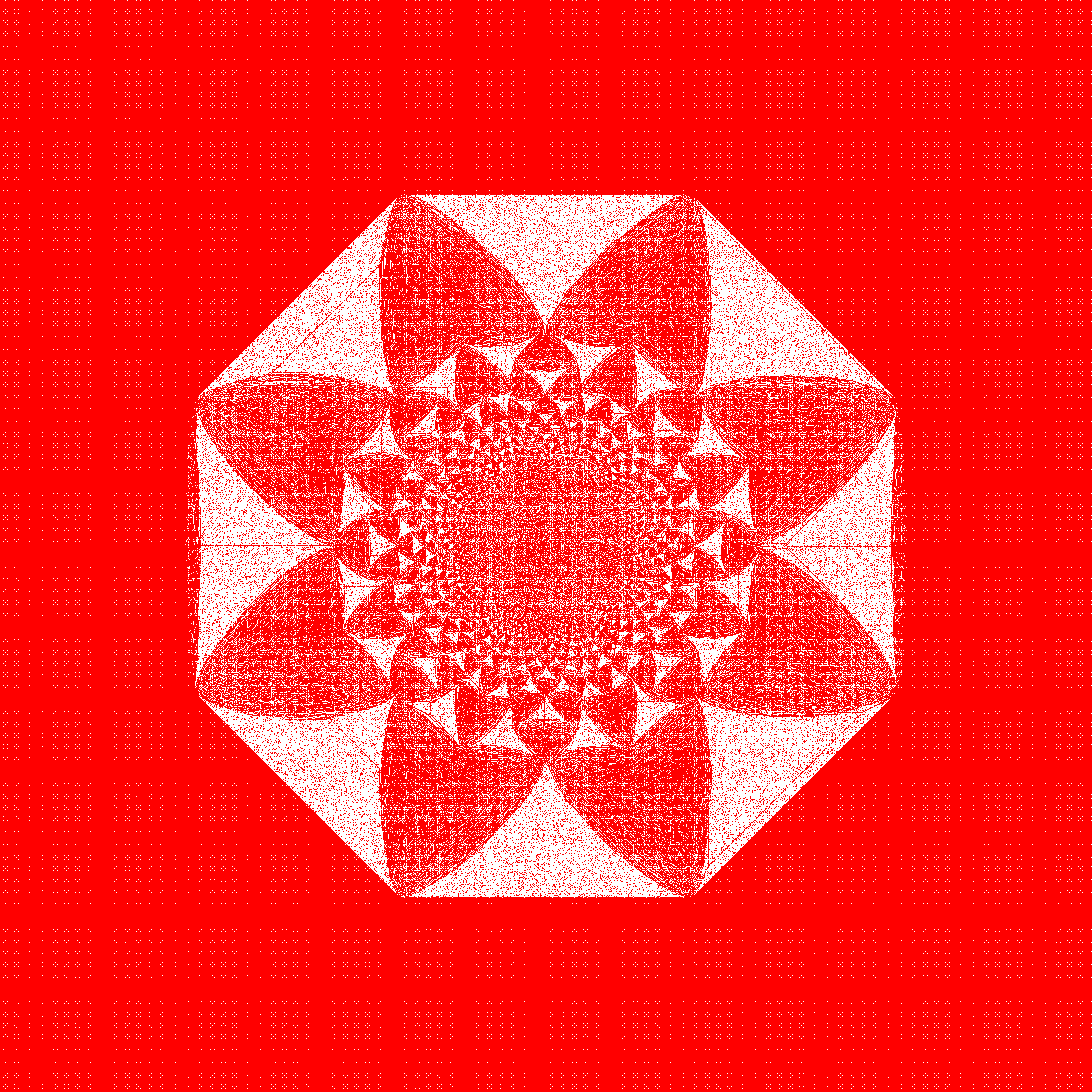}
\begin{picture}(0,0)
	\put(-370,150){\color{white}$(a)$}
	\put(-170,150){\color{white}$(b)$}
\end{picture}
\caption{Pattern grown on the F-lattice with noisy checkerboard background where
some of the heights $1$ are replaced by $0$'s. (a) $1\%$ sites changed, with
$N=228\times10^3$, and (b) $10\%$ changed, with $N= 896\times10^3$. Color code:
$0=$ red and $1=$ white. Zoom in for details in the electronic version.
Reproduced from \cite{tridib4}.
\label{fig:octagon_noise1} }
	\end{center}
\end{figure}
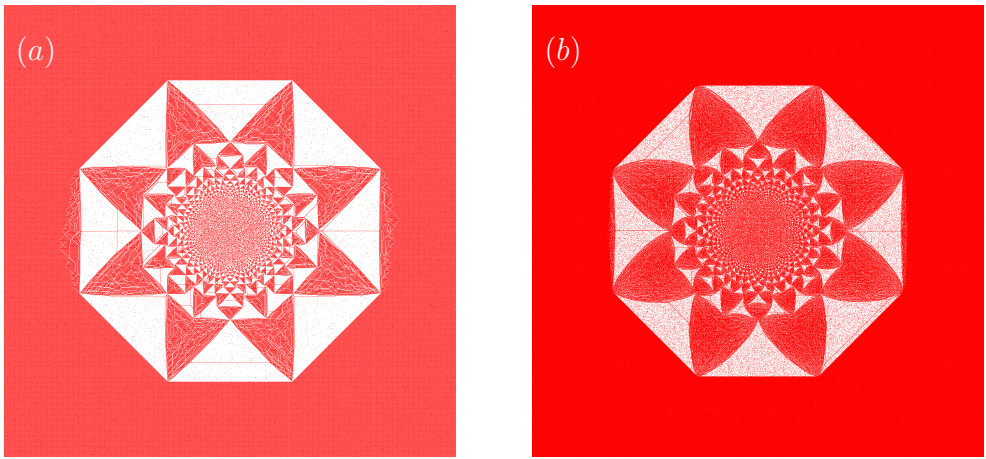
\begin{figure}
	\begin{center}
\includegraphics[width=6cm]{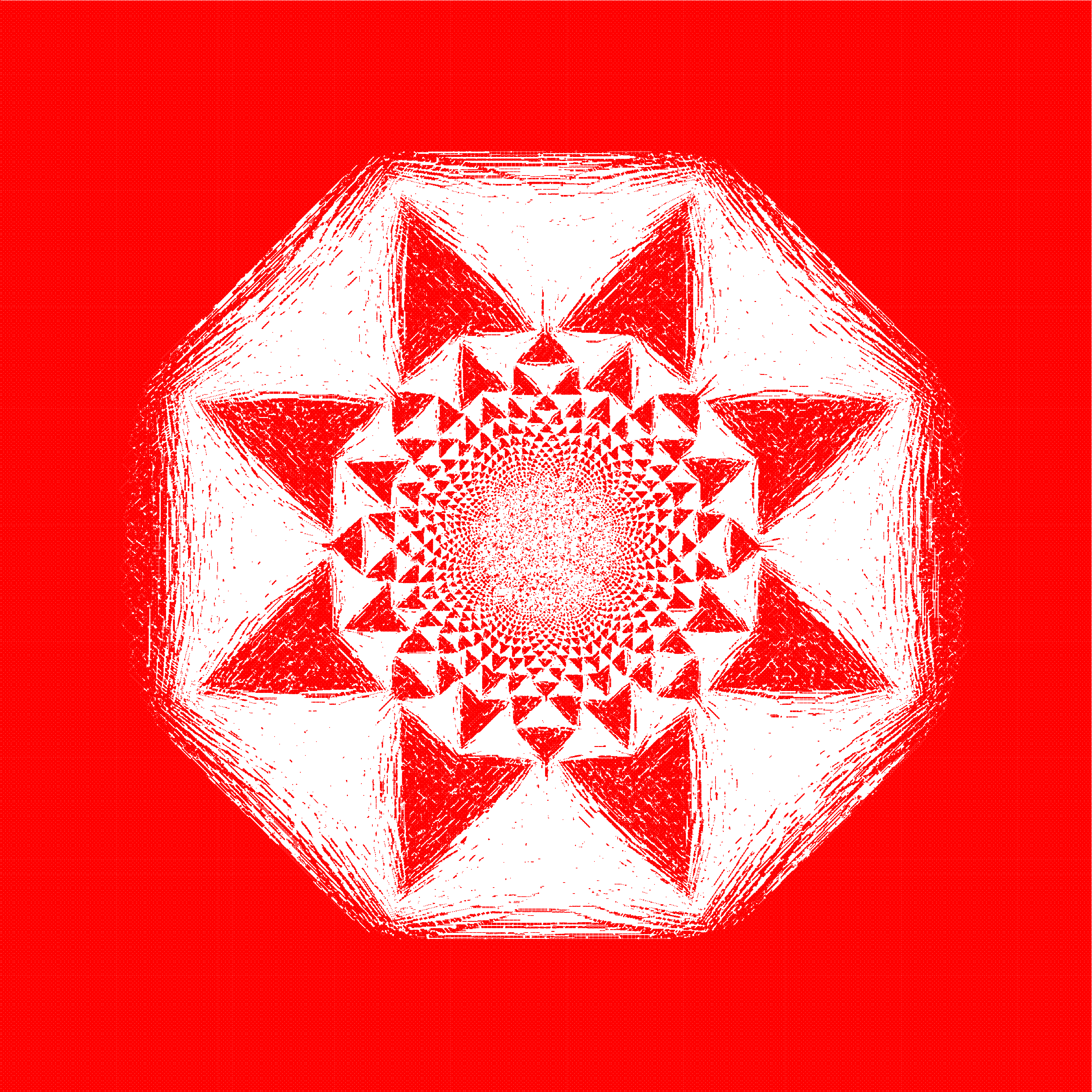}
\qquad
\includegraphics[width=6cm]{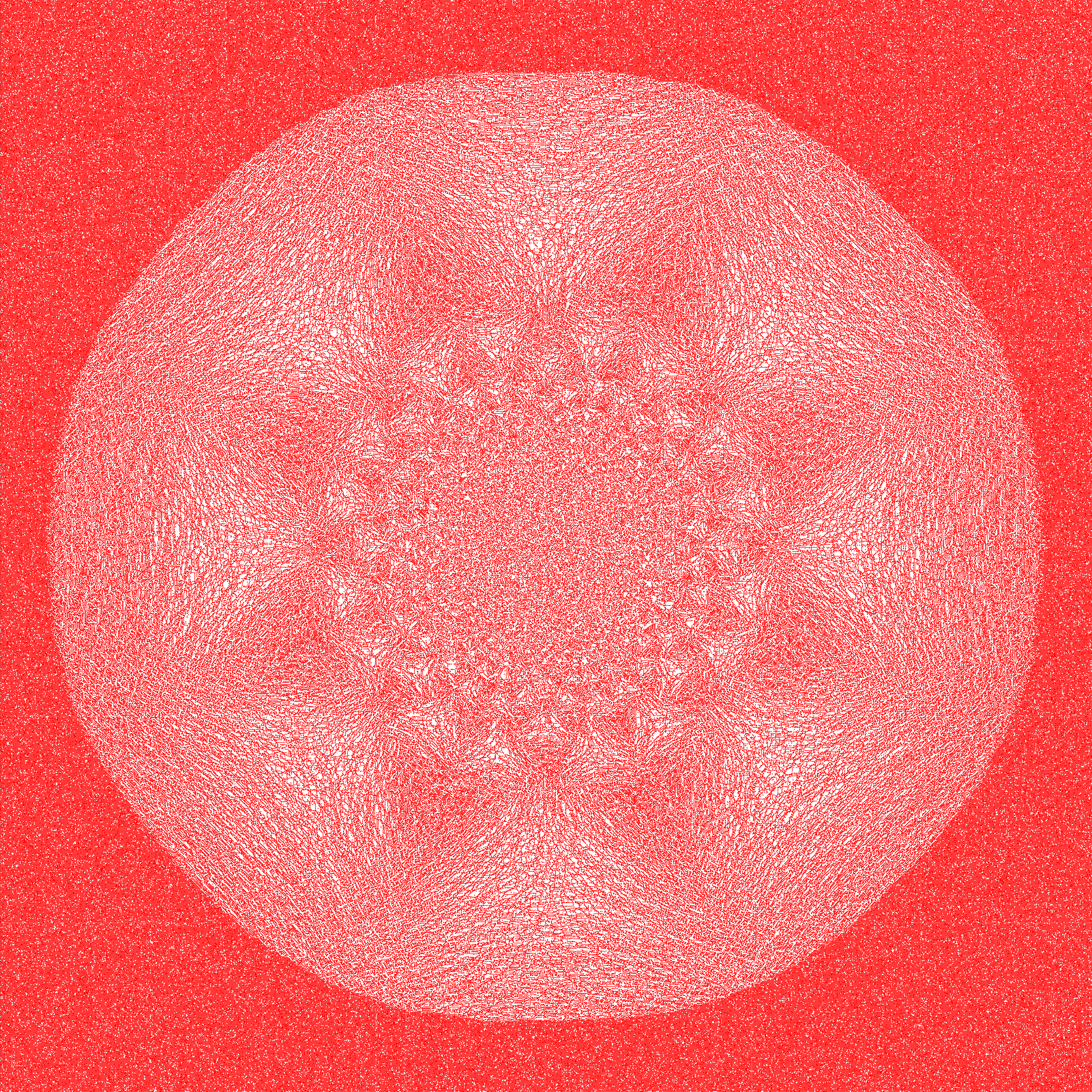}
\begin{picture}(0,0)
	\put(-370,150){\color{white}$(a)$}
	\put(-170,150){\color{white}$(b)$}
\end{picture}
\caption{Pattern generated on the F-lattice with noisy checkerboard background
where some heights are flipped. (a) $1\%$ sites
changed (b) $10\%$ sites changed. For both patterns $N=57\times 10^3$. Color code:
$0=$ red and $1=$ white. Zoom in for details in the electronic version.
Reproduced from \cite{tridib4}. \label{fig:octagon_noise2}}
	\end{center}
\end{figure}
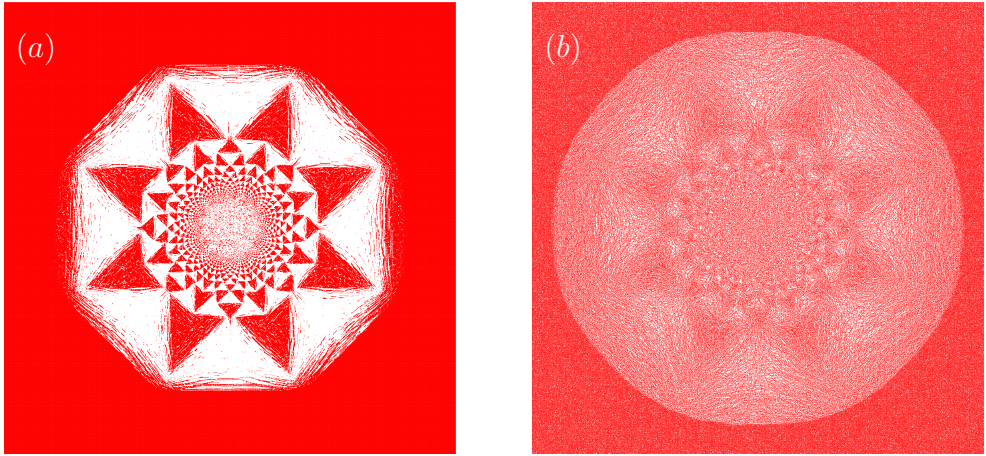

We also studied the problem when the initial starting  background pattern is not
a perfectly periodic pattern.  For the octagonal pattern on the F-lattice, we
studied the case when the initial configuration is the checkerboard pattern, with  a
small fraction of $1$'s randomly replaced by zeroes. The resulting patterns
are shown in \fref{fig:octagon_noise1} for two different values of the noise
strength. We find that one still gets a nontrivial pattern showing proportionate
growth. For small noise-strength, this pattern appears to be a small deformation
of the pattern without noise. The existence of sharp boundaries is quite
surprising,
as in the presence of noise, the potential function is no longer  piece-wise
quadratic. Since one can still define distinct patches in this case, we can also form their adjacency graph, which does not  depend on
the noise strength \cite{tridib4}.

If we take the F-Lattice checkerboard background as the starting point, and to
add noise, we flip the height of a small fraction of sites at random,  changing
height $1$ to zero, and also change $0$ to $1$, the resulting pattern is shown in
\fref{fig:octagon_noise2}, for two different strengths of the noise. We
still see proportionate growth, with the asymptotic pattern showing  the basic
structure as the pattern without noise. However, now, there are no sharp
boundaries between patches of low- and high-densities. There is an inhomogeneous
density profile  of particles in the asymptotic pattern, and this profile grows
proportionately. The amplitude of the ripple in the density pattern decreases
as the noise strength is increased.

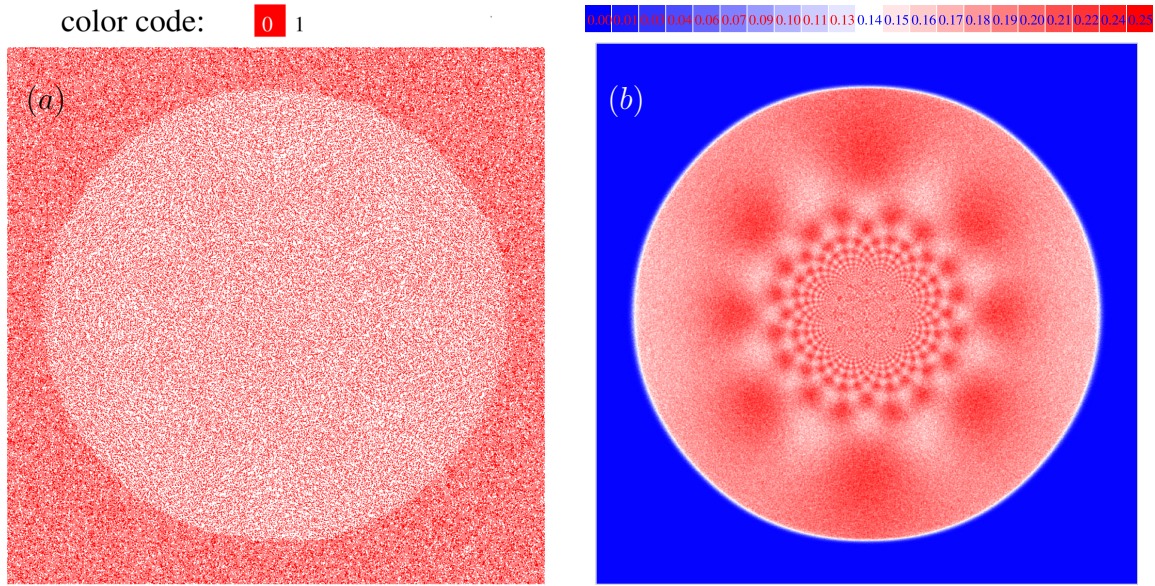
\begin{figure}
	\begin{center}
\includegraphics[width=0.455\textwidth]{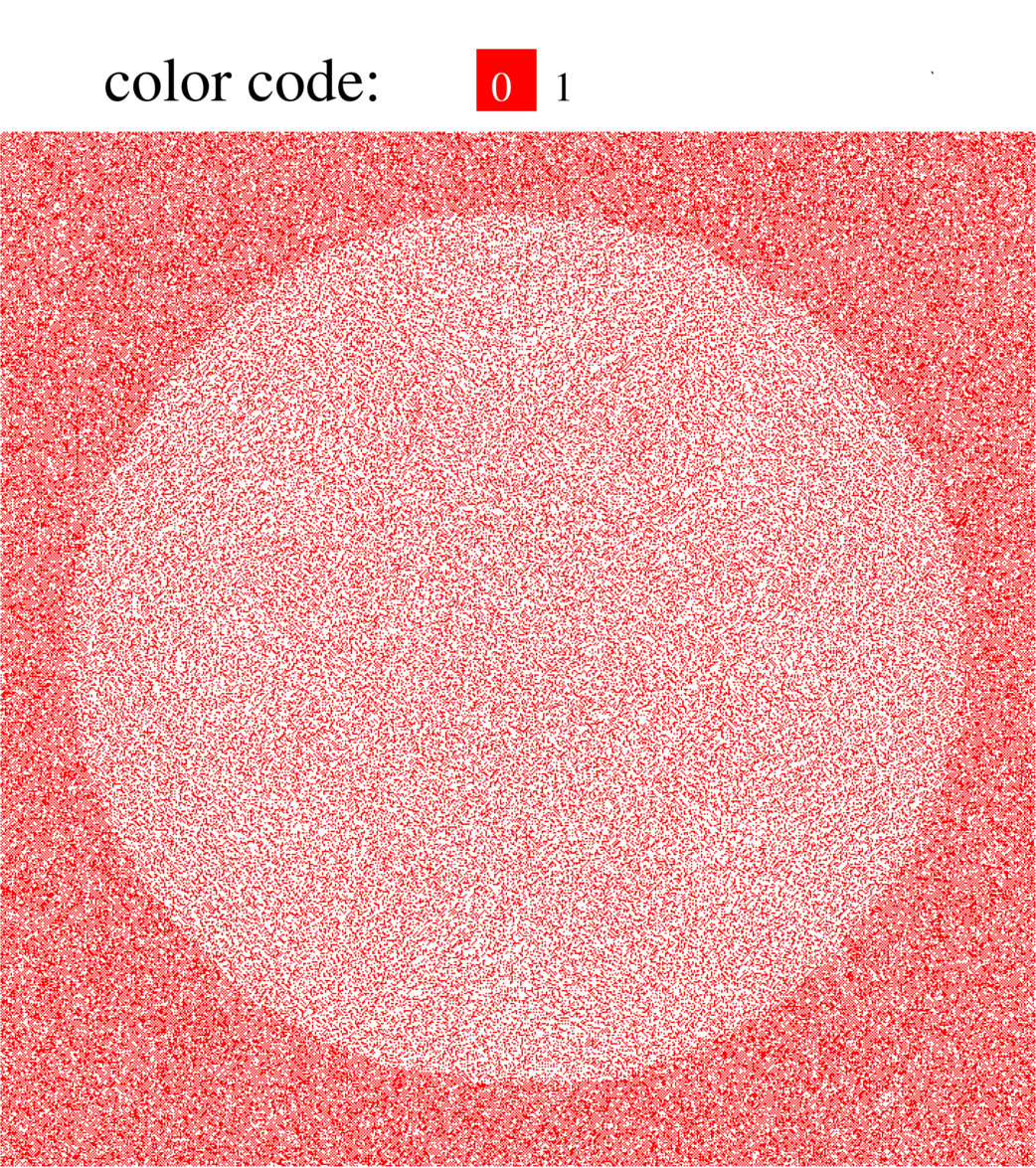}
\quad
\includegraphics[width=0.48\textwidth]{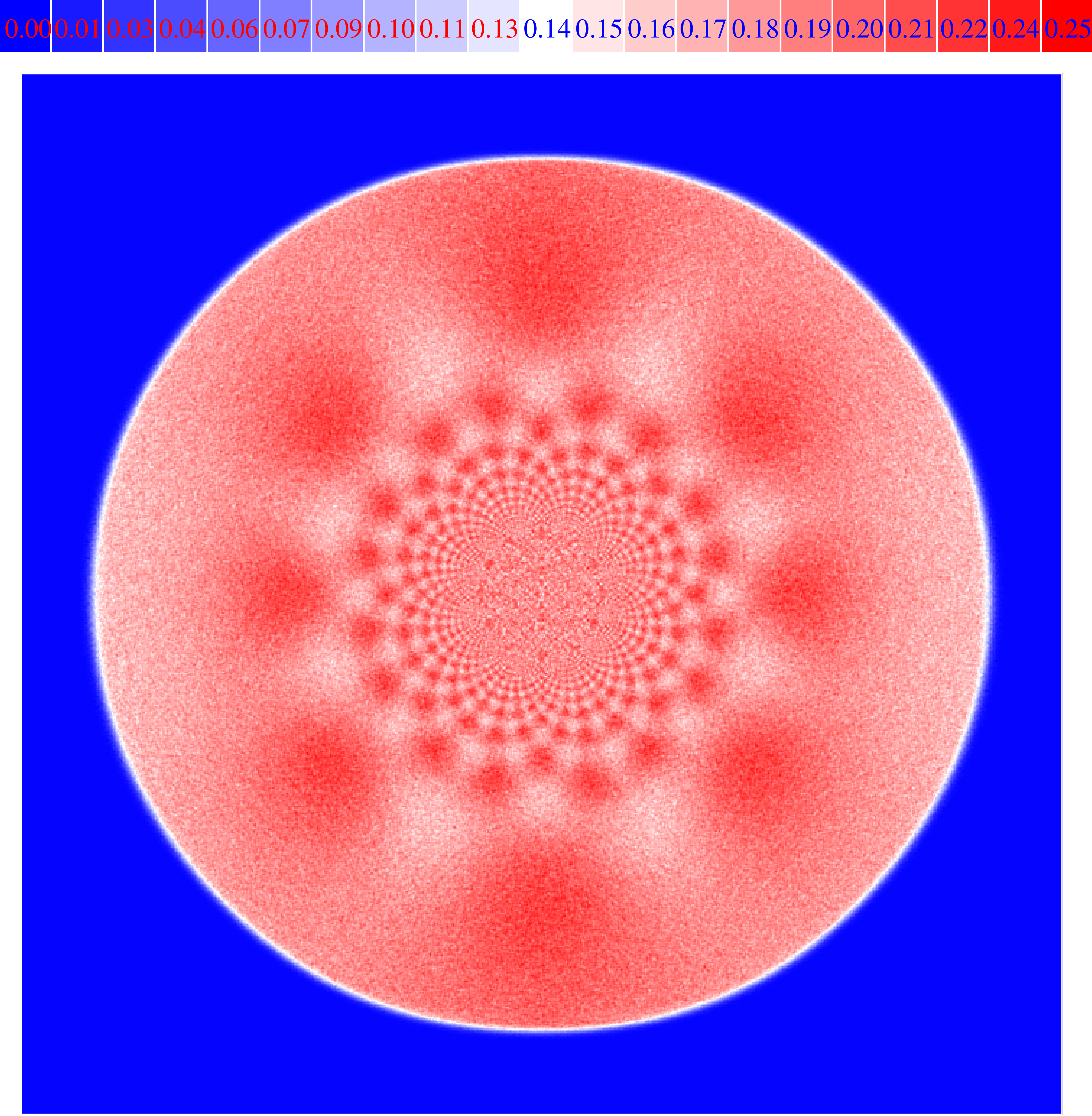}
\begin{picture}(0,0)
	\put(-430,180){$(a)$}
	\put(-210,180){$\color{white}(b)$}
\end{picture}
\caption{Pattern by adding $N=57\times10^3$ particles on F-lattice, checkerboard
background with $20\%$ sites flipped. (a) Pattern corresponding to a single
realization of the noisy background. (b) The pattern in terms of difference in
height from the background, averaged over $10^5$ realizations and also over the
sublattice. Zoom in for details in the electronic
version.
\label{fig:octagon_noise3}}
	\end{center}
\end{figure}

At higher noise levels, the density-inhomogeneity in  the pattern is not
immediately visible with naked eyes, as may be seen in Fig.
\ref{fig:octagon_noise3}. Here in \fref{fig:octagon_noise3}(a), we have shown
the pattern formed with one particular realization of noise. However, if one
forms several such patterns (with different realizations of noise in the
initial background), and determines the mean change in density of particles, one can
clearly see the nontrivial density profile in the pattern [ Fig. 11 (b)]. 

The very weak density inhomogeneity is reminiscent of convection patterns in
the Raleigh-Benard problem, near the onset of convection.  However, the origin
of instability in the density profile  in this problem is not yet understood.

\section{Other issues \label{sec:anisotropy}}
If we allow dissipation in the topplings, then the growth tends to saturate.
It is easy to see that the growth does not saturate exactly, and for
deterministic toppling rules with dissipation, the diameter increases  as $\log
N$ \cite{dandekar}. If the dissipation is stochastic, and there is a small
probability $\epsilon$ that a particle is lost per toppling, then the diameter
still increases, but only as $\log N$ for $N \gg 1/\epsilon$. Also, while noise
in toppling rules seems to wipe out  the intricate details of the pattern
without noise, some large scale features of the pattern, \textit{e.g.} the non-circular
outer boundary with some straight line-segments, seem to survive for $N\gg
1/\epsilon$. \cite{dandekar}

\begin{figure}
	\centering
	\includegraphics[width=0.6\textwidth,trim=1.5 0 0 0,clip]{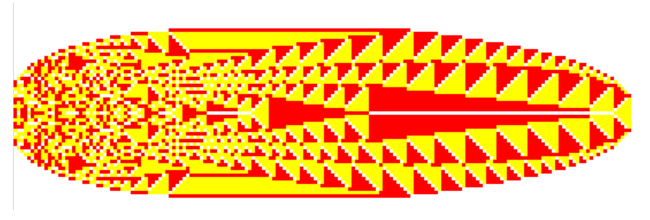}
	\caption{The `larva' pattern formed with partially directed transfer
	rules on a square lattice with $N=10^4$ grains with initial background of
	all heights being zero. The color code used: $0=$ white, $1=$ red and
	$2=$ yellow. \label{fig:bug2d}}
\end{figure}

One can also study directed particle transfer rules. For example, we take a
square lattice, with heights $\geq 3$ unstable, and particles are transferred in
only along the north, south and east directions. The corresponding pattern is shown
in \fref{fig:bug2d}. The basic unit of organization is not a periodic patch but
a sliver made of aligned squares of slowly varying width. The pattern does not
have proportionate growth, as different directions grow at different rate. Here,
the transverse size along the $y$-axis grows as $\xi_{\perp} \sim N^{1/3}$,
while the size along the growth direction $\xi_{\parallel} \sim N^{2/3}$.

\begin{figure}
	\begin{center}
\includegraphics[width=0.9\textwidth]{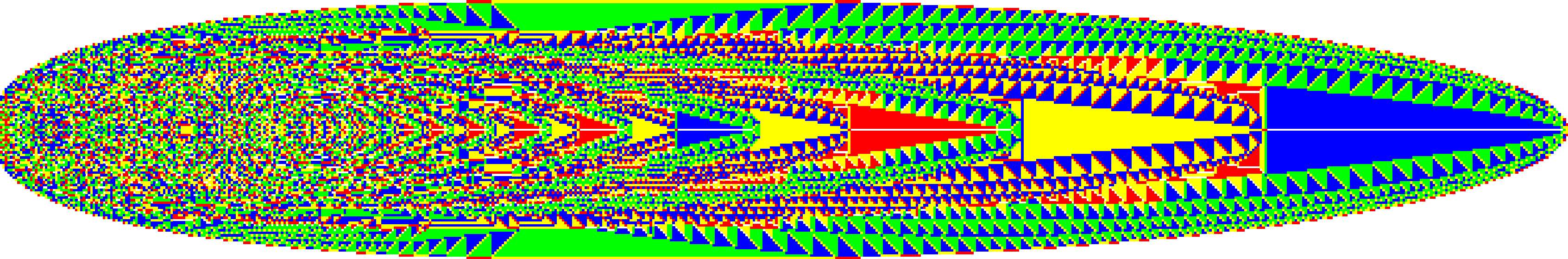}
\begin{picture}(0,0)
	\put(-410,60){$(a)$}
	\setlength{\unitlength}
	{0.8cm}
	\multiput(-15.5,-0.5)(0,0.4){10}{\line(0,1){0.2}}
	\multiput(-11.92,-0.5)(0,0.4){10}{\line(0,1){0.2}}
	\multiput(-3.43,-0.5)(0,0.4){10}{\line(0,1){0.2}}
\end{picture}
\vskip2ex
	\subfloat{
	\begin{minipage}[c][1\width]{0.33\textwidth}
		\centering
\includegraphics[width=1.0\textwidth,trim=8 8 8 8,clip]{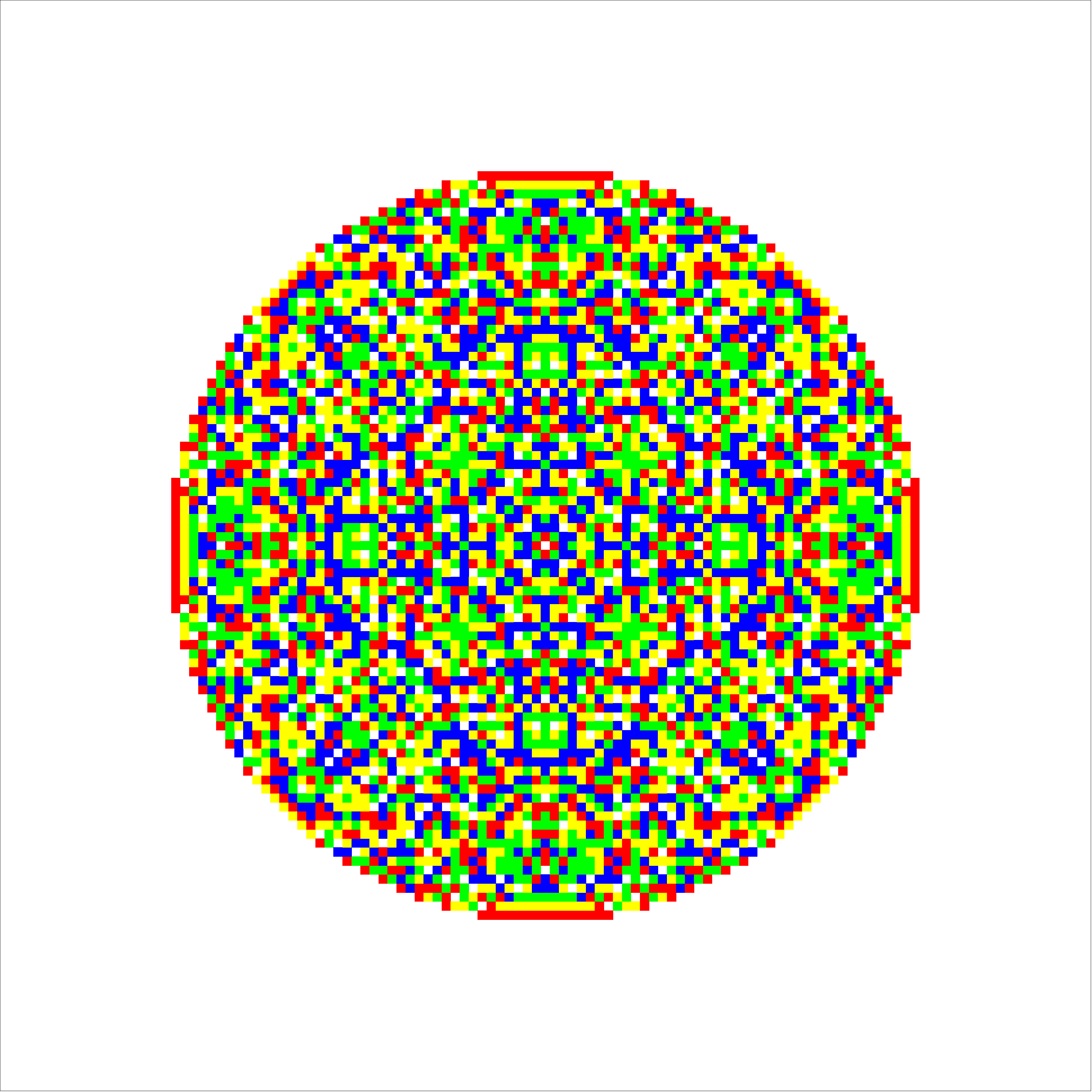}
\begin{picture}(0,0)
	\put(-70,120){$(b)$}
\end{picture}
	\end{minipage}}
	\subfloat{
	\begin{minipage}[c][1\width]{0.33\textwidth}
		\centering
\includegraphics[width=1.0\textwidth,trim=8 8 8 8,clip]{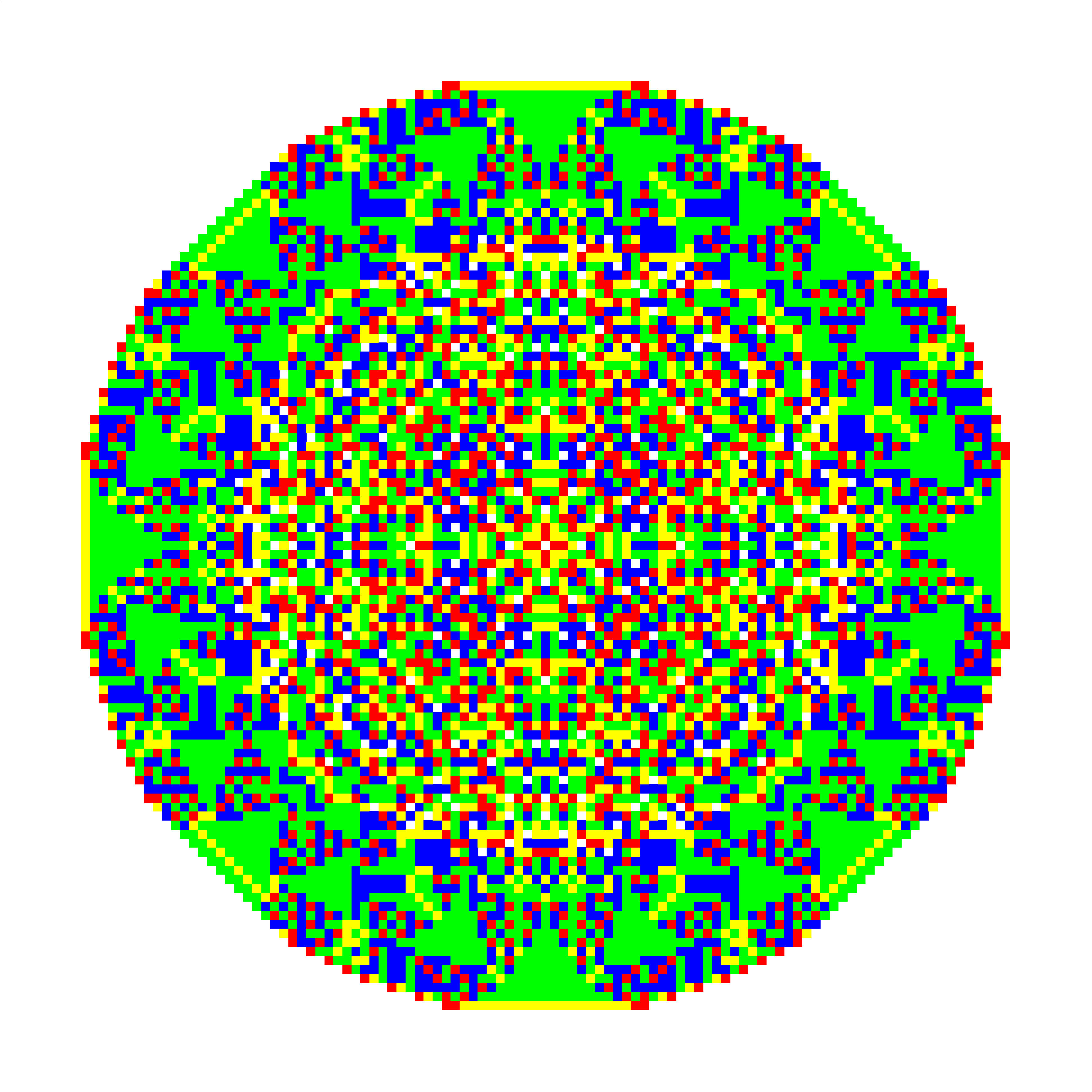}
\begin{picture}(0,0)
	\put(-80,120){$(c)$}
\end{picture}
	\end{minipage}}
	\subfloat{
	\begin{minipage}[c][1\width]{0.33\textwidth}
		\centering
\includegraphics[width=1.0\textwidth,trim=8 8 8 8,clip]{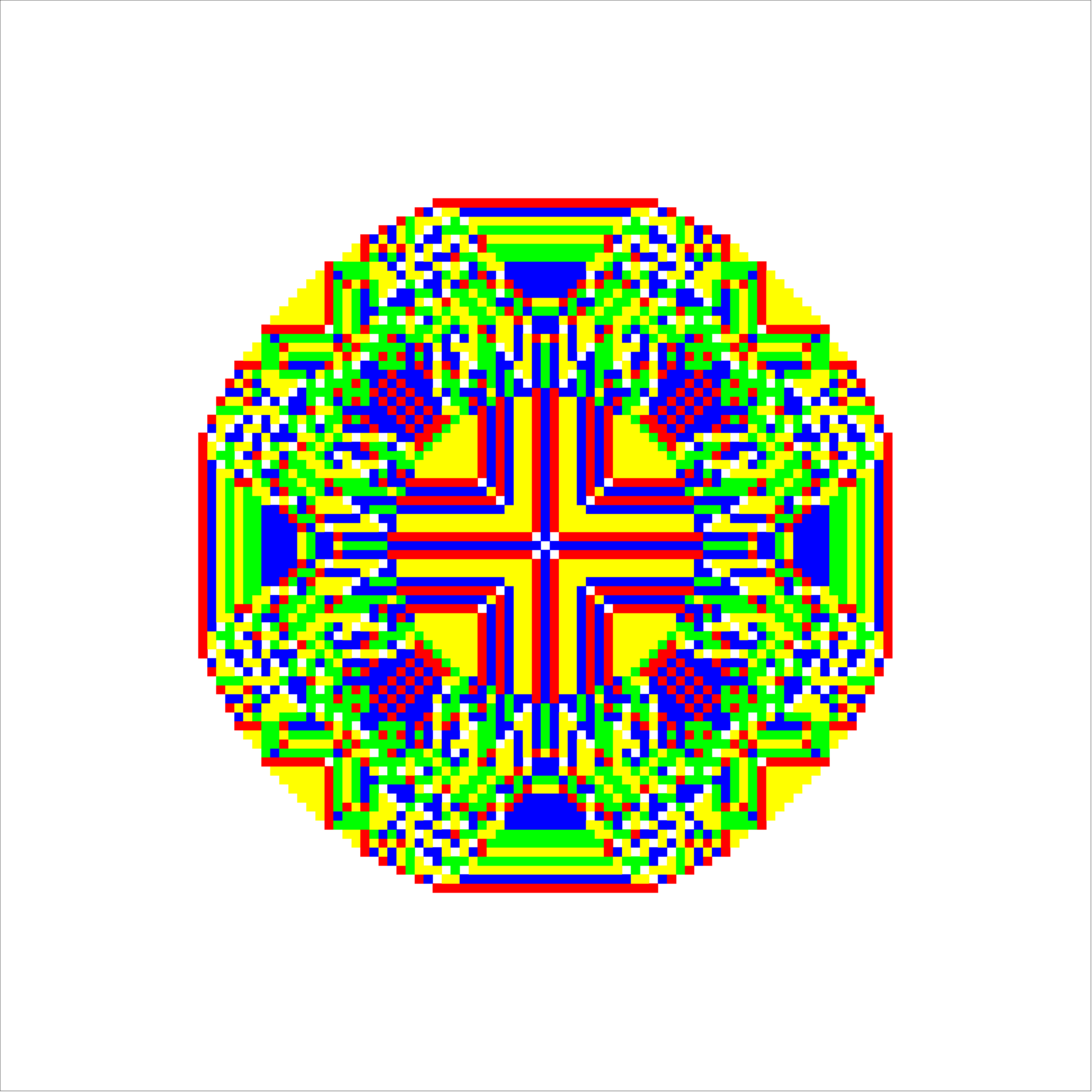}
\begin{picture}(0,0)
	\put(-70,120){$(d)$}
\end{picture}
	\end{minipage}}
\caption{A three-dimensional `larva' pattern formed on a directed cubic lattice.
The pattern was obtained by adding $N=10^7$ grains on an initial background of all heights being zero. The particles were added at the central point of the left end.  (a) shows the pattern
on a plane dissecting the three dimensional pattern along the longitudinal
direction. Here the colour code is $0=$ white, $1=$ red,
$2=$ yellow, $3=$ blue and $4=$ green. The particles are added at central point the left end.
(b)-(d) show cross sections perpendicular to the body axis
at different positions indicated by the dotted lines in (a). \label{fig:bug}}
\end{center}
\end{figure}
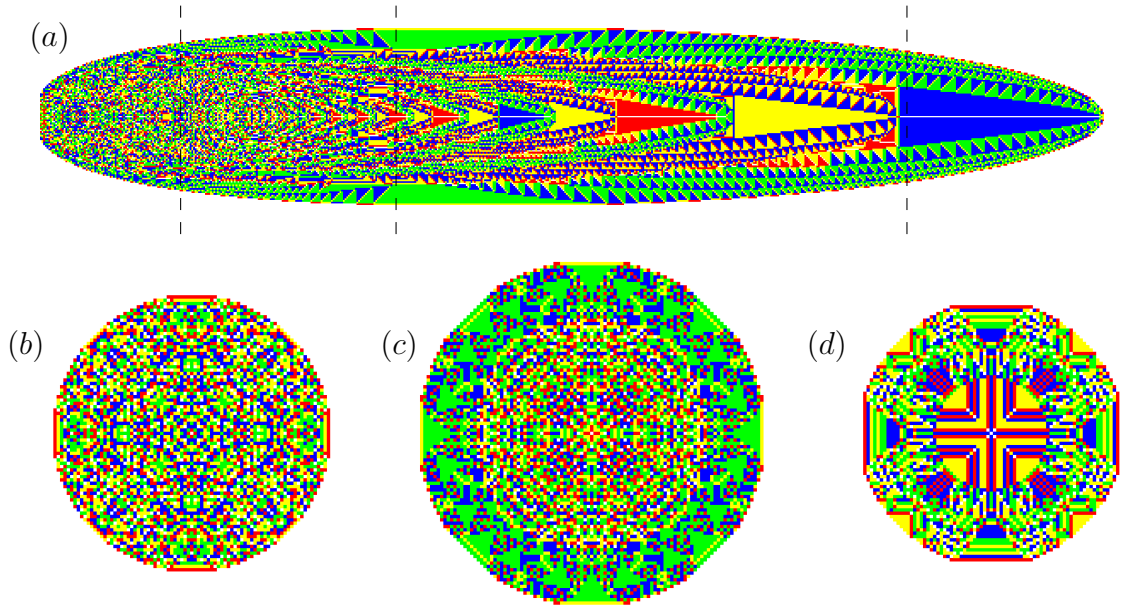

In three dimensions, the pattern formed by similar directed transfer rules on a
cubic lattice is shown in \fref{fig:bug}. Here the critical threshold is $5$, and on
toppling, one particle is transferred along each neighbor except in the negative
z-direction. We see that, in this pattern also, three  distinctive
head-thorax-abdomen type of features can be identified.

Because of the existence of two length scales, the arrangement of patches in
these systems shows a more complicated structure. The typical size of a patch
does not scale linearly with the diameter, and the fractional area of a patch in
the asymptotic pattern is zero.  Detailed characterization  of such patterns has
not been undertaken so far.

\section{Concluding remarks \label{sec:conclusion}}
To summarize,  we have argued that growing sandpile model
generates interesting  complex patterns, and shows proportionate growth using
very simple evolution rules.  They thus provide a simple mathematical model of
this fascinating phenomena.

While we have mainly emphasized the property of proportionate growth, we would like to also note their importance for understanding
morphogenesis. The classical model of morphogenesis is the Turing instability
\cite{review1}. In this, one gets only a limited set of basic patterns with a
small number of chemicals. Using  the large choice of toppling rules, and
starting backgrounds possible, the number of possible patterns that can be
generated in abelian sandpiles is unlimited. 

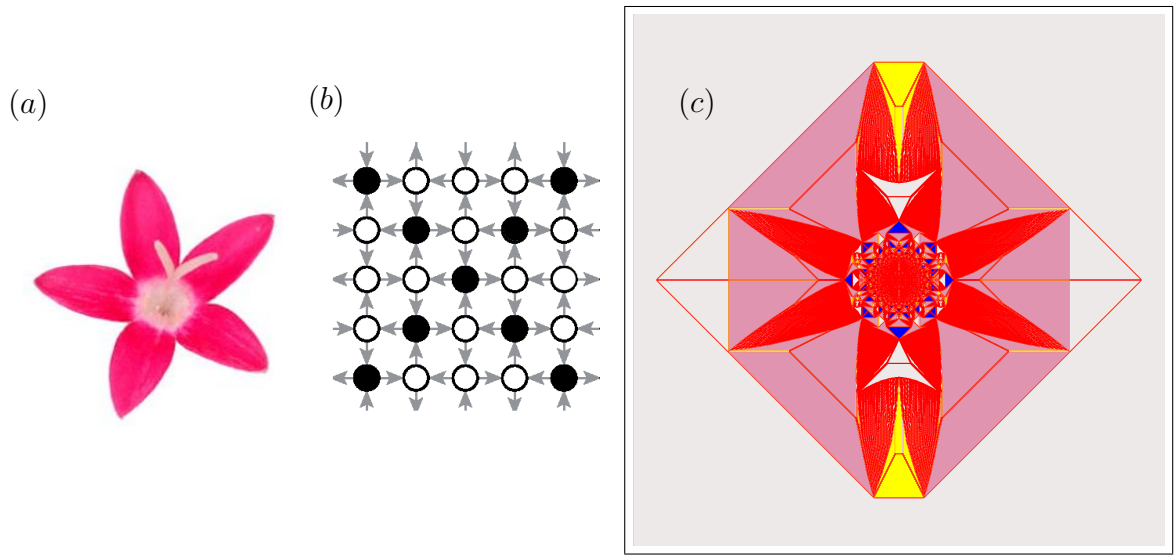
\begin{figure}
	\subfloat{
	\begin{minipage}[c][1\width]{0.25\textwidth}
		\centering
\includegraphics[width=0.9\textwidth]{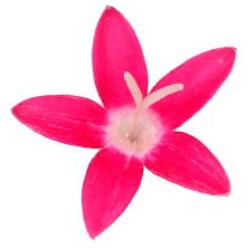}
\begin{picture}(0,0)
	\put(-110,120){$(a)$}
\end{picture}
	\end{minipage}}
	\subfloat{
	\begin{minipage}[c][1\width]{0.25\textwidth}
		\centering
\includegraphics[width=1\textwidth]{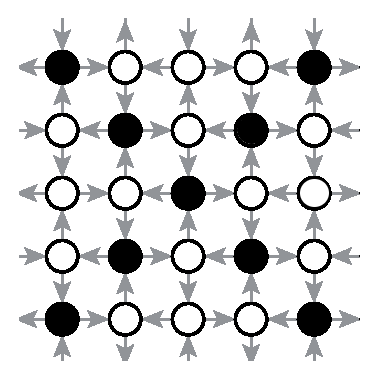}
\begin{picture}(0,0)
	\put(-60,135){$(b)$}
\end{picture}
	\end{minipage}}
	\subfloat{
	\begin{minipage}[c][1\width]{0.45\textwidth}
		\centering
\fbox{\includegraphics[width=1\textwidth]{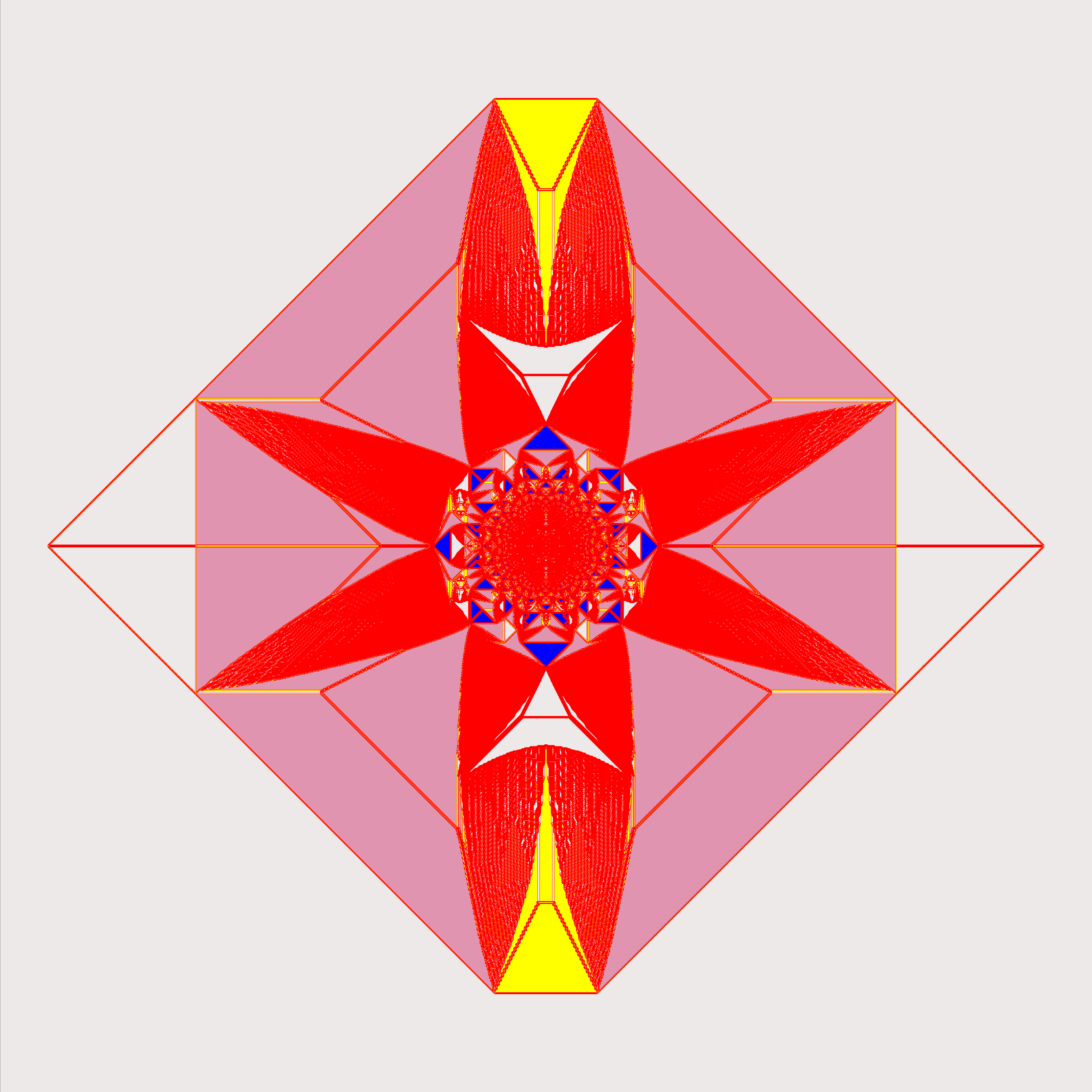}}
\begin{picture}(0,0)
	\put(-80,180){$(c)$}
\end{picture}
	\end{minipage}}
\caption{(a) A flower. (b) Unit cell of the background pattern with tilted squares on the
F-lattice. The unfilled circles denote height $1$, whereas the filled ones
represent empty sites. (c) The pattern produced by adding $N=256\times10^3$ particles on the
background. Different colours denote densities of particles, averaged over the unit cell of the
background pattern. Zoom in for details in the electronic version. \label{fig:corolla}}
\end{figure}

The most intriguing  feature of
these patterns is that  these minimal models can
sometimes produce patterns that have a striking similarity to the natural
ones. For example, the patterns in \fref{fig:bug2d} and \fref{fig:bug} looks like a larva, with parts that look
like the head, thorax, and the abdomen. There is no indication of these features
in the toppling rules defining the pattern. In \fref{fig:corolla}, we show
a pattern generated  on the F-lattice, using a background with unit cell made
of tilted squares of side-length $\ell$. By varying $\ell$, we can get a
flower-like pattern, and the petals of the flower become longer for larger
$\ell$.  The pattern shown was generated for $\ell =4$. One may have guessed
the pattern would have  bilateral symmetry, and  petal-like structures, but the
appearance of the anthers- and  circular corolla-  like structures is totally
unexpected.  Clearly, the model of sandpile toppling rules studied here is
rather simplistic, but  it is able to capture some crucial
elements of the real, much more complicated phenomena.

Our  motivations for our study of these patterns have been not only  their
beauty and fascinating diversity, but also the fact that they are
analytically tractable. Our present understanding of this pattern-formation
aspect of the problem is only partial. One has to take some important features
of the patterns as experimentally seen. It is hard to derive these
theoretically, directly from the definition of the problem. For example, in
the F-lattice octagonal pattern of \fref{fig:octagon}, we can determine the
exact metric properties of the resulting pattern, by solving the Laplace's
equation on  the adjacency graph of the pattern. But we have not been able to
deduce the structure the adjacency graph, starting  from the toppling rules.

There are many more complex patterns, for which the exact characterization
seems  more difficult, and would perhaps require some new  approach.  For
fast-growing sandpiles, with the exponent $a=1$, it seems plausible that the
mathematical  techniques of tropical  algebra \cite{tropical} could be useful.
Also, for the fractal patterns like the one shown in \fref{fig:bat}, the
calculation of the fractal dimension is a challenging problem.  There is an
intriguing  connection of the patterns generated here with the problem of
Apollonian packing of circles \cite{appollonian}. The `pattern selection
problem' of predicting which patterns will be found for a given  background, and
given toppling rules is not understood at present. The variational formulation
of this problem in terms of the `principle of least action' seems to be a promising
direction for further work \cite{tridib4}. Also, much more needs to be done to
understand the pattern formation on noisy backgrounds.

\ack
Our work has benefitted from discussions and collaboration  with a large number of colleagues.  DD
would like to thank  S. Chandra, R. Dandekar, J. P. Eckmann, A. Libchaber, V.
Nanjundiah, S. Nagel, S. Ostojic, L. Levine, K. Mopari, and S.B. Singha. DD's work has
been supported in part by a J. C. Bose fellowship by the government of India.

\section*{References}
\bibliography{reference}
\bibliographystyle{unsrt}
%
%
%
%
%
%
%
%
%
%
%
%
%

\end{document}